\pdfoutput=1
\documentclass[aps, prd, reprint,
superscriptaddress, amsmath, amssymb, floatfix,
showpacs, subfigure]{revtex4-2}

\usepackage[utf8]{inputenc}
\usepackage{graphicx}
\usepackage{dcolumn}

\usepackage[colorlinks, urlcolor=blue, citecolor=., menucolor=., anchorcolor=., linkcolor=., runcolor=.]{hyperref} 
\usepackage{orcidlink}
\usepackage{bm}
\usepackage{lineno}
\usepackage{float}
\usepackage{xspace}
\usepackage[inline]{enumitem}
\usepackage{overpic}
\usepackage{xcolor}
\usepackage{placeins}
\usepackage{subfigure}
\usepackage{booktabs}
\usepackage{longtable}
\usepackage[nowatermark]{fixmetodonotes}
\usepackage{tabulary}
\usepackage{isotope}
\usepackage[T1]{fontenc}
\usepackage{multirow}
\usepackage[utf8]{inputenc}
\usepackage{graphicx}
\usepackage{dcolumn}

\usepackage{bm}
\usepackage{placeins}
\newenvironment{conditions*} 
  {\par\vspace{\abovedisplayskip}\noindent
   \tabularx{\columnwidth}{>{$}l<{$} @{${}={}$} >{\raggedright\arraybackslash}X}}
  {\endtabularx\par\vspace{\belowdisplayskip}}

\usepackage{array}
\newcolumntype{x}[1]{>{\centering\arraybackslash\hspace{0pt}}p{#1}}

\usepackage{array}
\newcolumntype{L}[1]{>{\raggedright\let\newline\\\arraybackslash\hspace{0pt}}p{#1}}
\newcolumntype{C}[1]{>{\centering\let\newline\\\arraybackslash\hspace{0pt}}p{#1}}
\newcolumntype{R}[1]{>{\raggedleft\let\newline\\\arraybackslash\hspace{0pt}}p{#1}}

\usepackage{lineno}

\begin{document}
\title{First Measurement of the Total Inelastic Cross-Section of Positively-Charged Kaons on Argon at Energies Between 5.0 and 7.5 GeV}
\collaboration{The DUNE Collaboration}
\noaffiliation
%

\newcommand{\Abilene}{Abilene Christian University, Abilene, TX 79601, USA}
\newcommand{\Albanysuny}{University of Albany, SUNY, Albany, NY 12222, USA}
\newcommand{\Amsterdam}{University of Amsterdam, NL-1098 XG Amsterdam, The Netherlands}
\newcommand{\Antalya}{Antalya Bilim University, 07190 D\"o{\c s}emealtı/Antalya, Turkey}
\newcommand{\Antananarivo}{University of Antananarivo, Antananarivo 101, Madagascar}
\newcommand{\Antioquia}{University of Antioquia, Medell\'in, Colombia}
\newcommand{\AntonioNarino}{Universidad Antonio Nari\~no, Bogot\'a, Colombia}
\newcommand{\Argonne}{Argonne National Laboratory, Argonne, IL 60439, USA}
\newcommand{\Arizona}{University of Arizona, Tucson, AZ 85721, USA}
\newcommand{\Asuncion}{Universidad Nacional de Asunci\'on, San Lorenzo, Paraguay}
\newcommand{\Athens}{University of Athens, Zografou GR 157 84, Greece}
\newcommand{\Atlantico}{Universidad del Atl\'antico, Puerto Colombia, Atl\'antico, Colombia}
\newcommand{\Augustana}{Augustana University, Sioux Falls, SD 57197, USA}
\newcommand{\Bern}{University of Bern, CH-3012 Bern, Switzerland}
\newcommand{\Beykent}{Beykent University, Istanbul, Turkey}
\newcommand{\Birmingham}{University of Birmingham, Birmingham B15 2TT, United Kingdom}
\newcommand{\BolognaUniversity}{Universit\`a di Bologna, 40127 Bologna, Italy}
\newcommand{\Boston}{Boston University, Boston, MA 02215, USA}
\newcommand{\Bristol}{University of Bristol, Bristol BS8 1TL, United Kingdom}
\newcommand{\Brookhaven}{Brookhaven National Laboratory, Upton, NY 11973, USA}
\newcommand{\Bucharest}{University of Bucharest, Bucharest, Romania}
\newcommand{\CalBerkeley}{University of California Berkeley, Berkeley, CA 94720, USA}
\newcommand{\CalDavis}{University of California Davis, Davis, CA 95616, USA}
\newcommand{\CalIrvine}{University of California Irvine, Irvine, CA 92697, USA}
\newcommand{\CalLosangeles}{University of California Los Angeles, Los Angeles, CA 90095, USA}
\newcommand{\CalRiverside}{University of California Riverside, Riverside CA 92521, USA}
\newcommand{\CalSantabarbara}{University of California Santa Barbara, Santa Barbara, CA 93106, USA}
\newcommand{\Caltech}{California Institute of Technology, Pasadena, CA 91125, USA}
\newcommand{\Cambridge}{University of Cambridge, Cambridge CB3 0HE, United Kingdom}
\newcommand{\Campinas}{Universidade Estadual de Campinas, Campinas - SP, 13083-970, Brazil}
\newcommand{\CataniaUniversitadi}{Universit\`a di Catania, 2 - 95131 Catania, Italy}
\newcommand{\Catolica}{Universidad Cat\'olica del Norte, Antofagasta, Chile}
\newcommand{\CBPF}{Centro Brasileiro de Pesquisas F\'isicas, Rio de Janeiro, RJ 22290-180, Brazil}
\newcommand{\CEASaclay}{IRFU, CEA, Universit\'e Paris-Saclay, F-91191 Gif-sur-Yvette, France}
\newcommand{\CERN}{CERN, The European Organization for Nuclear Research, 1211 Meyrin, Switzerland}
\newcommand{\Charles}{Institute of Particle and Nuclear Physics of the Faculty of Mathematics and Physics of the Charles University, 180 00 Prague 8, Czech Republic }
\newcommand{\Chicago}{University of Chicago, Chicago, IL 60637, USA}
\newcommand{\ChungAng}{Chung-Ang University, Seoul 06974, South Korea}
\newcommand{\CIEMAT}{CIEMAT, Centro de Investigaciones Energ\'eticas, Medioambientales y Tecnol\'ogicas, E-28040 Madrid, Spain}
\newcommand{\Cincinnati}{University of Cincinnati, Cincinnati, OH 45221, USA}
\newcommand{\Cinvestav}{Centro de Investigaci\'on y de Estudios Avanzados del Instituto Polit\'ecnico Nacional (Cinvestav), Mexico City, Mexico}
\newcommand{\Colima}{Universidad de Colima, Colima, Mexico}
\newcommand{\ColoradoBoulder}{University of Colorado Boulder, Boulder, CO 80309, USA}
\newcommand{\ColoradoState}{Colorado State University, Fort Collins, CO 80523, USA}
\newcommand{\Columbia}{Columbia University, New York, NY 10027, USA}
\newcommand{\conida}{Comisi\'on Nacional de Investigaci\'on y Desarrollo Aeroespacial, Lima, Peru}
\newcommand{\Cti}{Centro de Tecnologia da Informacao Renato Archer, Amarais - Campinas, SP - CEP 13069-901}
\newcommand{\CUSB}{Central University of South Bihar, Gaya, 824236, India
}
\newcommand{\CzechAcademyofSciences}{Institute of Physics, Czech Academy of Sciences, 182 00 Prague 8, Czech Republic}
\newcommand{\CzechTechnical}{Czech Technical University, 115 19 Prague 1, Czech Republic}
\newcommand{\DannecyleVieux}{Laboratoire d'Annecy de Physique des Particules, Universit\'e Savoie Mont Blanc, CNRS, LAPP-IN2P3, 74000 Annecy, France}
\newcommand{\Daresbury}{Daresbury Laboratory, Cheshire WA4 4AD, United Kingdom}
\newcommand{\Dordt}{Dordt University, Sioux Center, IA 51250, USA}
\newcommand{\Drexel}{Drexel University, Philadelphia, PA 19104, USA}
\newcommand{\Duke}{Duke University, Durham, NC 27708, USA}
\newcommand{\Durham}{Durham University, Durham DH1 3LE, United Kingdom}
\newcommand{\Edinburgh}{University of Edinburgh, Edinburgh EH8 9YL, United Kingdom}
\newcommand{\EIA}{Universidad EIA, Envigado, Antioquia, Colombia}
\newcommand{\Eotvos}{E\"otv\"os Lor\'and University, 1053 Budapest, Hungary}
\newcommand{\erciyes}{Erciyes University, Kayseri, Turkey}
\newcommand{\FCULport}{Faculdade de Ci\^encias da Universidade de Lisboa - FCUL, 1749-016 Lisboa, Portugal}
\newcommand{\FederaldeAlfenas}{Universidade Federal de Alfenas, Po{\c c}os de Caldas - MG, 37715-400, Brazil}
\newcommand{\FederaldeGoias}{Universidade Federal de Goias, Goiania, GO 74690-900, Brazil}
\newcommand{\FederaldoABC}{Universidade Federal do ABC, Santo Andr\'e - SP, 09210-580, Brazil}
\newcommand{\FederaldoRio}{Universidade Federal do Rio de Janeiro, Rio de Janeiro - RJ, 21941-901, Brazil}
\newcommand{\Fermi}{Fermi National Accelerator Laboratory, Batavia, IL 60510, USA}
\newcommand{\Ferrarauniv}{University of Ferrara, Ferrara, Italy}
\newcommand{\Florida}{University of Florida, Gainesville, FL 32611-8440, USA}
\newcommand{\Floridastate}{Florida State University, Tallahassee, FL, 32306 USA}
\newcommand{\Fluminense}{Fluminense Federal University, 9 Icara\'i Niter\'oi - RJ, 24220-900, Brazil }
\newcommand{\Genova}{Universit\`a degli Studi di Genova, Genova, Italy}
\newcommand{\Georgian}{Georgian Technical University, Tbilisi, Georgia}
\newcommand{\Granada}{University of Granada \& CAFPE, 18002 Granada, Spain}
\newcommand{\GranSasso}{Gran Sasso Science Institute, L'Aquila, Italy}
\newcommand{\GranSassoLab}{Laboratori Nazionali del Gran Sasso, L'Aquila AQ, Italy}
\newcommand{\Grenoble}{University Grenoble Alpes, CNRS, Grenoble INP, LPSC-IN2P3, 38000 Grenoble, France}
\newcommand{\Guanajuato}{Universidad de Guanajuato, Guanajuato, C.P. 37000, Mexico}
\newcommand{\Harish}{Harish-Chandra Research Institute, Jhunsi, Allahabad 211 019, India}
\newcommand{\Hawaii}{University of Hawaii, Honolulu, HI 96822, USA}
\newcommand{\hkust}{Hong Kong University of Science and Technology, Kowloon, Hong Kong, China}
\newcommand{\Houston}{University of Houston, Houston, TX 77204, USA}
\newcommand{\Hyderabad}{University of  Hyderabad, Gachibowli, Hyderabad - 500 046, India}
\newcommand{\Idaho}{Idaho State University, Pocatello, ID 83209, USA}
\newcommand{\IFIC}{Instituto de F\'isica Corpuscular, CSIC and Universitat de Val\`encia, 46980 Paterna, Valencia, Spain}
\newcommand{\IGFAE}{Instituto Galego de F\'isica de Altas Enerx\'ias, University of Santiago de Compostela, Santiago de Compostela, 15782, Spain}
\newcommand{\Iitk}{Indian Institute of Technology Kanpur, Uttar Pradesh 208016, India}
\newcommand{\Illinoisinstitute}{Illinois Institute of Technology, Chicago, IL 60616, USA}
\newcommand{\Imperial}{Imperial College of Science Technology and Medicine, London SW7 2BZ, United Kingdom}
\newcommand{\IndGuwahati}{Indian Institute of Technology Guwahati, Guwahati, 781 039, India}
\newcommand{\IndHyderabad}{Indian Institute of Technology Hyderabad, Hyderabad, 502285, India}
\newcommand{\Indiana}{Indiana University, Bloomington, IN 47405, USA}
\newcommand{\INFNBologna}{Istituto Nazionale di Fisica Nucleare Sezione di Bologna, 40127 Bologna BO, Italy}
\newcommand{\INFNCatania}{Istituto Nazionale di Fisica Nucleare Sezione di Catania, I-95123 Catania, Italy}
\newcommand{\INFNFerrara}{Istituto Nazionale di Fisica Nucleare Sezione di Ferrara, I-44122 Ferrara, Italy}
\newcommand{\INFNFrascati}{Istituto Nazionale di Fisica Nucleare Laboratori Nazionali di Frascati, Frascati, Roma, Italy}
\newcommand{\INFNGenova}{Istituto Nazionale di Fisica Nucleare Sezione di Genova, 16146 Genova GE, Italy}
\newcommand{\INFNLecce}{Istituto Nazionale di Fisica Nucleare Sezione di Lecce, 73100 - Lecce, Italy}
\newcommand{\INFNMilanBicocca}{Istituto Nazionale di Fisica Nucleare Sezione di Milano Bicocca, 3 - I-20126 Milano, Italy}
\newcommand{\INFNMilano}{Istituto Nazionale di Fisica Nucleare Sezione di Milano, 20133 Milano, Italy}
\newcommand{\INFNNapoli}{Istituto Nazionale di Fisica Nucleare Sezione di Napoli, I-80126 Napoli, Italy}
\newcommand{\INFNPadova}{Istituto Nazionale di Fisica Nucleare Sezione di Padova, 35131 Padova, Italy}
\newcommand{\INFNPavia}{Istituto Nazionale di Fisica Nucleare Sezione di Pavia,  I-27100 Pavia, Italy}
\newcommand{\INFNPisa}{Istituto Nazionale di Fisica Nucleare Laboratori Nazionali di Pisa, Pisa PI, Italy}
\newcommand{\INFNRoma}{Istituto Nazionale di Fisica Nucleare Sezione di Roma, 00185 Roma RM, Italy}
\newcommand{\INFNSud}{Istituto Nazionale di Fisica Nucleare Laboratori Nazionali del Sud, 95123 Catania, Italy}
\newcommand{\Ingenieria}{Universidad Nacional de Ingenier\'ia, Lima 25, Per\'u}
\newcommand{\Insubria }{University of Insubria, Via Ravasi, 2, 21100 Varese VA, Italy}
\newcommand{\Iowa}{University of Iowa, Iowa City, IA 52242, USA}
\newcommand{\IowaState}{Iowa State University, Ames, Iowa 50011, USA}
\newcommand{\IPLyon}{Institut de Physique des 2 Infinis de Lyon, 69622 Villeurbanne, France}
\newcommand{\IPM}{Institute for Research in Fundamental Sciences, Tehran, Iran}
\newcommand{\ISTlisboa}{Instituto Superior T\'ecnico - IST, Universidade de Lisboa, 1049-001 Lisboa, Portugal}
\newcommand{\Ita}{Instituto Tecnol\'ogico de Aeron\'autica, Sao Jose dos Campos, Brazil}
\newcommand{\Iwate}{Iwate University, Morioka, Iwate 020-8551, Japan}
\newcommand{\Jacksonstate}{Jackson State University, Jackson, MS 39217, USA}
\newcommand{\Jawaharlal}{Jawaharlal Nehru University, New Delhi 110067, India}
\newcommand{\Jeonbuk}{Jeonbuk National University, Jeonrabuk-do 54896, South Korea}
\newcommand{\Jyvaskyla}{Jyv\"askyl\"a University, FI-40014 Jyv\"askyl\"a, Finland}
\newcommand{\Kansasstate}{Kansas State University, Manhattan, KS 66506, USA}
\newcommand{\Kavli}{Kavli Institute for the Physics and Mathematics of the Universe, Kashiwa, Chiba 277-8583, Japan}
\newcommand{\KEK}{High Energy Accelerator Research Organization (KEK), Ibaraki, 305-0801, Japan}
\newcommand{\KISTI}{Korea Institute of Science and Technology Information, Daejeon, 34141, South Korea}
\newcommand{\Kure}{National Institute of Technology, Kure College, Hiroshima, 737-8506, Japan}
\newcommand{\Kyiv}{Taras Shevchenko National University of Kyiv, 01601 Kyiv, Ukraine}
\newcommand{\Lancaster}{Lancaster University, Lancaster LA1 4YB, United Kingdom}
\newcommand{\LawrenceBerkeley}{Lawrence Berkeley National Laboratory, Berkeley, CA 94720, USA}
\newcommand{\LIP}{Laborat\'orio de Instrumenta{\c c}\~ao e F\'isica Experimental de Part\'iculas, 1649-003 Lisboa and 3004-516 Coimbra, Portugal}
\newcommand{\Liverpool}{University of Liverpool, L69 7ZE, Liverpool, United Kingdom}
\newcommand{\LosAlmos}{Los Alamos National Laboratory, Los Alamos, NM 87545, USA}
\newcommand{\Louisanastate}{Louisiana State University, Baton Rouge, LA 70803, USA}
\newcommand{\LpBordeaux}{Laboratoire de Physique des Deux Infinis Bordeaux - IN2P3, F-33175 Gradignan, Bordeaux, France, }
\newcommand{\Lucknow}{University of Lucknow, Uttar Pradesh 226007, India}
\newcommand{\Madrid}{Madrid Autonoma University and IFT UAM/CSIC, 28049 Madrid, Spain}
\newcommand{\Mainz}{Johannes Gutenberg-Universit\"at Mainz, 55122 Mainz, Germany}
\newcommand{\Manchester}{University of Manchester, Manchester M13 9PL, United Kingdom}
\newcommand{\Massinsttech}{Massachusetts Institute of Technology, Cambridge, MA 02139, USA}
\newcommand{\Medellin}{University of Medell\'in, Medell\'in, 050026 Colombia }
\newcommand{\Michigan}{University of Michigan, Ann Arbor, MI 48109, USA}
\newcommand{\Michiganstate}{Michigan State University, East Lansing, MI 48824, USA}
\newcommand{\MilanoBicocca}{Universit\`a di Milano Bicocca , 20126 Milano, Italy}
\newcommand{\MilanoUniv}{Universit\`a degli Studi di Milano, I-20133 Milano, Italy}
\newcommand{\Minnduluth}{University of Minnesota Duluth, Duluth, MN 55812, USA}
\newcommand{\Minntwin}{University of Minnesota Twin Cities, Minneapolis, MN 55455, USA}
\newcommand{\Mississippi}{University of Mississippi, University, MS 38677 USA}
\newcommand{\napoli}{Universit\`a degli Studi di Napoli Federico II , 80138 Napoli NA, Italy}
\newcommand{\Nikhef}{Nikhef National Institute of Subatomic Physics, 1098 XG Amsterdam, Netherlands}
\newcommand{\Niser}{National Institute of Science Education and Research (NISER), Odisha 752050, India}
\newcommand{\Northdakota}{University of North Dakota, Grand Forks, ND 58202-8357, USA}
\newcommand{\Northernillinois}{Northern Illinois University, DeKalb, IL 60115, USA}
\newcommand{\Northwestern}{Northwestern University, Evanston, Il 60208, USA}
\newcommand{\NotreDame}{University of Notre Dame, Notre Dame, IN 46556, USA}
\newcommand{\NoviSad}{University of Novi Sad, 21102 Novi Sad, Serbia}
\newcommand{\Occidental}{Occidental College, Los Angeles, CA  90041}
\newcommand{\Ohiostate}{Ohio State University, Columbus, OH 43210, USA}
\newcommand{\OregonState}{Oregon State University, Corvallis, OR 97331, USA}
\newcommand{\Oxford}{University of Oxford, Oxford, OX1 3RH, United Kingdom}
\newcommand{\PacificNorthwest}{Pacific Northwest National Laboratory, Richland, WA 99352, USA}
\newcommand{\Padova}{Universt\`a degli Studi di Padova, I-35131 Padova, Italy}
\newcommand{\Panjab}{Panjab University, Chandigarh, 160014, India}
\newcommand{\Parissaclay}{Universit\'e Paris-Saclay, CNRS/IN2P3, IJCLab, 91405 Orsay, France}
\newcommand{\Parisuniversite}{Universit\'e Paris Cit\'e, CNRS, Astroparticule et Cosmologie, Paris, France}
\newcommand{\Parma}{University of Parma,  43121 Parma PR, Italy}
\newcommand{\Pavia}{Universit\`a degli Studi di Pavia, 27100 Pavia PV, Italy}
\newcommand{\Penn}{University of Pennsylvania, Philadelphia, PA 19104, USA}
\newcommand{\PennState}{Pennsylvania State University, University Park, PA 16802, USA}
\newcommand{\PhysicalResearchLaboratory}{Physical Research Laboratory, Ahmedabad 380 009, India}
\newcommand{\Pisa}{Universit\`a di Pisa, I-56127 Pisa, Italy}
\newcommand{\Pitt}{University of Pittsburgh, Pittsburgh, PA 15260, USA}
\newcommand{\Pontificia}{Pontificia Universidad Cat\'olica del Per\'u, Lima, Per\'u}
\newcommand{\PuertoRico}{University of Puerto Rico, Mayaguez 00681, Puerto Rico, USA}
\newcommand{\Punjab}{Punjab Agricultural University, Ludhiana 141004, India}
\newcommand{\QMUL}{Queen Mary University of London, London E1 4NS, United Kingdom
}
\newcommand{\Radboud}{Radboud University, NL-6525 AJ Nijmegen, Netherlands}
\newcommand{\Rice}{Rice University, Houston, TX 77005}
\newcommand{\Rochester}{University of Rochester, Rochester, NY 14627, USA}
\newcommand{\Royalholloway}{Royal Holloway College London, London, TW20 0EX, United Kingdom}
\newcommand{\Rutgers}{Rutgers University, Piscataway, NJ, 08854, USA}
\newcommand{\Rutherford}{STFC Rutherford Appleton Laboratory, Didcot OX11 0QX, United Kingdom}
\newcommand{\Salento}{Universit\`a del Salento, 73100 Lecce, Italy}
\newcommand{\santamarta}{Universidad del Magdalena, Santa Marta - Colombia}
\newcommand{\Sapienza}{Sapienza University of Rome, 00185 Roma RM, Italy}
\newcommand{\SergioArboleda}{Universidad Sergio Arboleda, 11022 Bogot\'a, Colombia}
\newcommand{\Sheffield}{University of Sheffield, Sheffield S3 7RH, United Kingdom}
\newcommand{\SLAC}{SLAC National Accelerator Laboratory, Menlo Park, CA 94025, USA}
\newcommand{\Southcarolina}{University of South Carolina, Columbia, SC 29208, USA}
\newcommand{\SouthDakotaSchool}{South Dakota School of Mines and Technology, Rapid City, SD 57701, USA}
\newcommand{\SouthDakotaState}{South Dakota State University, Brookings, SD 57007, USA}
\newcommand{\SouthernMethodist}{Southern Methodist University, Dallas, TX 75275, USA}
\newcommand{\StonyBrook}{Stony Brook University, SUNY, Stony Brook, NY 11794, USA}
\newcommand{\SURF}{Sanford Underground Research Facility, Lead, SD, 57754, USA}
\newcommand{\Sussex}{University of Sussex, Brighton, BN1 9RH, United Kingdom}
\newcommand{\Syracuse}{Syracuse University, Syracuse, NY 13244, USA}
\newcommand{\Tecnologica }{Universidade Tecnol\'ogica Federal do Paran\'a, Curitiba, Brazil}
\newcommand{\TelAviv}{Tel Aviv University, Tel Aviv-Yafo, Israel}
\newcommand{\TexasAMcollege}{Texas A\&M University, College Station, Texas 77840}
\newcommand{\TexasAMcorpuscristi}{Texas A\&M University - Corpus Christi, Corpus Christi, TX 78412, USA}
\newcommand{\TexasArlington}{University of Texas at Arlington, Arlington, TX 76019, USA}
\newcommand{\Texasaustin}{University of Texas at Austin, Austin, TX 78712, USA}
\newcommand{\Toronto}{University of Toronto, Toronto, Ontario M5S 1A1, Canada}
\newcommand{\Tufts}{Tufts University, Medford, MA 02155, USA}
\newcommand{\Unifesp}{Universidade Federal de S\~ao Paulo, 09913-030, S\~ao Paulo, Brazil}
\newcommand{\UNIST}{Ulsan National Institute of Science and Technology, Ulsan 689-798, South Korea}
\newcommand{\UniversityCollegeLondon}{University College London, London, WC1E 6BT, United Kingdom}
\newcommand{\UNMSM}{Universidad Nacional Mayor de San Marcos, Lima, Peru}
\newcommand{\ValleyCity}{Valley City State University, Valley City, ND 58072, USA}
\newcommand{\Vigo}{University of Vigo, E- 36310 Vigo Spain}
\newcommand{\VirginiaTech}{Virginia Tech, Blacksburg, VA 24060, USA}
\newcommand{\Warsaw}{University of Warsaw, 02-093 Warsaw, Poland}
\newcommand{\Warwick}{University of Warwick, Coventry CV4 7AL, United Kingdom}
\newcommand{\Wellesley}{Wellesley College, Wellesley, MA 02481, USA}
\newcommand{\Wichita}{Wichita State University, Wichita, KS 67260, USA}
\newcommand{\WilliamMary}{William and Mary, Williamsburg, VA 23187, USA}
\newcommand{\Wisconsin}{University of Wisconsin Madison, Madison, WI 53706, USA}
\newcommand{\Yale}{Yale University, New Haven, CT 06520, USA}
\newcommand{\Yerevan}{Yerevan Institute for Theoretical Physics and Modeling, Yerevan 0036, Armenia}
\newcommand{\York}{York University, Toronto M3J 1P3, Canada}
\affiliation{\Abilene}
\affiliation{\Albanysuny}
\affiliation{\Amsterdam}
\affiliation{\Antalya}
\affiliation{\Antananarivo}
\affiliation{\Antioquia}
\affiliation{\AntonioNarino}
\affiliation{\Argonne}
\affiliation{\Arizona}
\affiliation{\Asuncion}
\affiliation{\Athens}
\affiliation{\Atlantico}
\affiliation{\Augustana}
\affiliation{\Bern}
\affiliation{\Beykent}
\affiliation{\Birmingham}
\affiliation{\BolognaUniversity}
\affiliation{\Boston}
\affiliation{\Bristol}
\affiliation{\Brookhaven}
\affiliation{\Bucharest}
\affiliation{\CalBerkeley}
\affiliation{\CalDavis}
\affiliation{\CalIrvine}
\affiliation{\CalLosangeles}
\affiliation{\CalRiverside}
\affiliation{\CalSantabarbara}
\affiliation{\Caltech}
\affiliation{\Cambridge}
\affiliation{\Campinas}
\affiliation{\CataniaUniversitadi}
\affiliation{\Catolica}
\affiliation{\CBPF}
\affiliation{\CEASaclay}
\affiliation{\CERN}
\affiliation{\Charles}
\affiliation{\Chicago}
\affiliation{\ChungAng}
\affiliation{\CIEMAT}
\affiliation{\Cincinnati}
\affiliation{\Cinvestav}
\affiliation{\Colima}
\affiliation{\ColoradoBoulder}
\affiliation{\ColoradoState}
\affiliation{\Columbia}
\affiliation{\conida}
\affiliation{\Cti}
\affiliation{\CUSB}
\affiliation{\CzechAcademyofSciences}
\affiliation{\CzechTechnical}
\affiliation{\DannecyleVieux}
\affiliation{\Daresbury}
\affiliation{\Dordt}
\affiliation{\Drexel}
\affiliation{\Duke}
\affiliation{\Durham}
\affiliation{\Edinburgh}
\affiliation{\EIA}
\affiliation{\Eotvos}
\affiliation{\erciyes}
\affiliation{\FCULport}
\affiliation{\FederaldeAlfenas}
\affiliation{\FederaldeGoias}
\affiliation{\FederaldoABC}
\affiliation{\FederaldoRio}
\affiliation{\Fermi}
\affiliation{\Ferrarauniv}
\affiliation{\Florida}
\affiliation{\Floridastate}
\affiliation{\Fluminense}
\affiliation{\Genova}
\affiliation{\Georgian}
\affiliation{\Granada}
\affiliation{\GranSasso}
\affiliation{\GranSassoLab}
\affiliation{\Grenoble}
\affiliation{\Guanajuato}
\affiliation{\Harish}
\affiliation{\Hawaii}
\affiliation{\hkust}
\affiliation{\Houston}
\affiliation{\Hyderabad}
\affiliation{\Idaho}
\affiliation{\IFIC}
\affiliation{\IGFAE}
\affiliation{\Iitk}
\affiliation{\Illinoisinstitute}
\affiliation{\Imperial}
\affiliation{\IndGuwahati}
\affiliation{\IndHyderabad}
\affiliation{\Indiana}
\affiliation{\INFNBologna}
\affiliation{\INFNCatania}
\affiliation{\INFNFerrara}
\affiliation{\INFNFrascati}
\affiliation{\INFNGenova}
\affiliation{\INFNLecce}
\affiliation{\INFNMilanBicocca}
\affiliation{\INFNMilano}
\affiliation{\INFNNapoli}
\affiliation{\INFNPadova}
\affiliation{\INFNPavia}
\affiliation{\INFNPisa}
\affiliation{\INFNRoma}
\affiliation{\INFNSud}
\affiliation{\Ingenieria}
\affiliation{\Insubria }
\affiliation{\Iowa}
\affiliation{\IowaState}
\affiliation{\IPLyon}
\affiliation{\IPM}
\affiliation{\ISTlisboa}
\affiliation{\Ita}
\affiliation{\Iwate}
\affiliation{\Jacksonstate}
\affiliation{\Jawaharlal}
\affiliation{\Jeonbuk}
\affiliation{\Jyvaskyla}
\affiliation{\Kansasstate}
\affiliation{\Kavli}
\affiliation{\KEK}
\affiliation{\KISTI}
\affiliation{\Kure}
\affiliation{\Kyiv}
\affiliation{\Lancaster}
\affiliation{\LawrenceBerkeley}
\affiliation{\LIP}
\affiliation{\Liverpool}
\affiliation{\LosAlmos}
\affiliation{\Louisanastate}
\affiliation{\LpBordeaux}
\affiliation{\Lucknow}
\affiliation{\Madrid}
\affiliation{\Mainz}
\affiliation{\Manchester}
\affiliation{\Massinsttech}
\affiliation{\Medellin}
\affiliation{\Michigan}
\affiliation{\Michiganstate}
\affiliation{\MilanoBicocca}
\affiliation{\MilanoUniv}
\affiliation{\Minnduluth}
\affiliation{\Minntwin}
\affiliation{\Mississippi}
\affiliation{\napoli}
\affiliation{\Nikhef}
\affiliation{\Niser}
\affiliation{\Northdakota}
\affiliation{\Northernillinois}
\affiliation{\Northwestern}
\affiliation{\NotreDame}
\affiliation{\NoviSad}
\affiliation{\Occidental}
\affiliation{\Ohiostate}
\affiliation{\OregonState}
\affiliation{\Oxford}
\affiliation{\PacificNorthwest}
\affiliation{\Padova}
\affiliation{\Panjab}
\affiliation{\Parissaclay}
\affiliation{\Parisuniversite}
\affiliation{\Parma}
\affiliation{\Pavia}
\affiliation{\Penn}
\affiliation{\PennState}
\affiliation{\PhysicalResearchLaboratory}
\affiliation{\Pisa}
\affiliation{\Pitt}
\affiliation{\Pontificia}
\affiliation{\PuertoRico}
\affiliation{\Punjab}
\affiliation{\QMUL}
\affiliation{\Radboud}
\affiliation{\Rice}
\affiliation{\Rochester}
\affiliation{\Royalholloway}
\affiliation{\Rutgers}
\affiliation{\Rutherford}
\affiliation{\Salento}
\affiliation{\santamarta}
\affiliation{\Sapienza}
\affiliation{\SergioArboleda}
\affiliation{\Sheffield}
\affiliation{\SLAC}
\affiliation{\Southcarolina}
\affiliation{\SouthDakotaSchool}
\affiliation{\SouthDakotaState}
\affiliation{\SouthernMethodist}
\affiliation{\StonyBrook}
\affiliation{\SURF}
\affiliation{\Sussex}
\affiliation{\Syracuse}
\affiliation{\Tecnologica }
\affiliation{\TelAviv}
\affiliation{\TexasAMcollege}
\affiliation{\TexasAMcorpuscristi}
\affiliation{\TexasArlington}
\affiliation{\Texasaustin}
\affiliation{\Toronto}
\affiliation{\Tufts}
\affiliation{\Unifesp}
\affiliation{\UNIST}
\affiliation{\UniversityCollegeLondon}
\affiliation{\UNMSM}
\affiliation{\ValleyCity}
\affiliation{\Vigo}
\affiliation{\VirginiaTech}
\affiliation{\Warsaw}
\affiliation{\Warwick}
\affiliation{\Wellesley}
\affiliation{\Wichita}
\affiliation{\WilliamMary}
\affiliation{\Wisconsin}
\affiliation{\Yale}
\affiliation{\Yerevan}
\affiliation{\York}
\author{A.~Abed Abud} \affiliation{\CERN}
\author{B.~Abi} \affiliation{\Oxford}
\author{R.~Acciarri} \affiliation{\Fermi}
\author{M.~A.~Acero} \affiliation{\Atlantico}
\author{M.~R.~Adames} \affiliation{\Tecnologica }
\author{G.~Adamov} \affiliation{\Georgian}
\author{M.~Adamowski} \affiliation{\Fermi}
\author{D.~Adams} \affiliation{\Brookhaven}
\author{M.~Adinolfi} \affiliation{\Bristol}
\author{C.~Adriano} \affiliation{\Campinas}
% ...
\collaboration{The DUNE Collaboration}
\noaffiliation

\begin{abstract}
    ProtoDUNE Single-Phase (ProtoDUNE-SP) is a 770-ton liquid argon time projection chamber that operated in a hadron test beam at the CERN Neutrino Platform in 2018. We present a measurement of the total inelastic cross section of charged kaons on argon as a function of kaon energy using 6 and 7 GeV/$c$ beam momentum settings. The flux-weighted average of the extracted inelastic cross section at each beam momentum setting was measured to be 380$\pm$26 mbarns for the 6 GeV/$c$ setting and 379$\pm$35 mbarns for the 7 GeV/$c$ setting. 
    \end{abstract}

\date{August 1st, 2024}

\maketitle

\section{Introduction}
\label{sec:intro}

Liquid argon time projection chambers (LArTPCs) may be used to measure the trajectories of charged particles with millimeter resolution. This capability makes the detectors, like those of the Deep Underground Neutrino Experiment (DUNE) far detector modules, sensitive to studying GeV-scale and MeV-scale neutrinos and searching for physics beyond the Standard Model~\cite{Abi_2020}. An example of important physics that can be done using the DUNE far detector modules is a search for proton decay to a final state with a neutrino and a charged kaon ($p\rightarrow \nu+K^+$), which is predicted to be dominant in a broad class of supersymmetric Grand Unified Theories~\cite{Sakai:1981pk,Weinberg:1981wj,Dimopoulos:1981dw,Dimopoulos:1981zb,Nath:1985ub}. Unlike searches in Water Cherenkov detectors~\cite{Super-Kamiokande:2014otb}, DUNE can detect the final-state kaon, which has a momentum of 330 MeV/$c$ absent final-state interactions. The efficiency of observing this signature is sensitive to modeling kaon transport in the LAr medium, which is limited by the dearth of kaon-argon scattering data. This search for nucleon decay requires a representative model of kaon transport and interactions in liquid argon to ensure an accurate simulation of signal events. Without reliable data and simulations, the relevant uncertainties for the kaon cross section on argon cannot be constrained. This can lead to large systematic uncertainties in nucleon decay searches with a potentially biased cross-section model.

As a first step toward collecting high-quality kaon-argon interaction data, the ProtoDUNE Single-Phase (ProtoDUNE-SP) large-scale prototype of a DUNE far detector module was exposed to a test beam from the H4-VLE beamline at CERN that included kaons at 6 and 7 GeV/$c$~\cite{abi2020first,Abed_Abud_2022}. ProtoDUNE-SP is a 770-ton LArTPC with the same drift distance and full-scale engineering parts as a DUNE Far Detector Horizontal Drift module. The detector has two drift volumes and has dimensions of 7.2 m wide, 6.1 m high, and 7 m long. It measures the tracking and calorimetry of charged particles by detecting the ionization electrons that drift toward three layers of wire planes. The H4-VLE beamline, a tertiary beam from the CERN Super Proton Synchrotron, is referred to as simply the ``beam'' in many places in this paper. ProtoDUNE-SP collected data from the beam, using many beamline momentum settings, over two months from September 2018 to November 2018.

The data from ProtoDUNE-SP can be used to inform predictions from event generators that simulate hadron-nucleus interactions, like the neutrino event generator GENIE~\cite{Andreopoulos:2009rq,Andreopoulos:2015wxa, Tena-Vidal:2021rpu, GENIE:2021wox, Dytman} and the transport and interaction simulation program \textsc{Geant4} ~\cite{GEANT4, geant4IEEE, ALLISON2016186}, and thus improve the modeling of kaon interactions on argon nuclei. The cross section has never been measured as a function of energy on argon. Therefore, the purpose of this analysis is to provide the first measurement of the total inelastic cross section of kaons on argon at these high energies. Neither GENIE nor \textsc{Geant4} has recommended uncertainties for kaon-argon interactions, providing a unique opportunity for ProtoDUNE-SP to inform inputs on associated modeling uncertainties.

\begin{figure}
    \centering
    \includegraphics[width=0.45\textwidth]{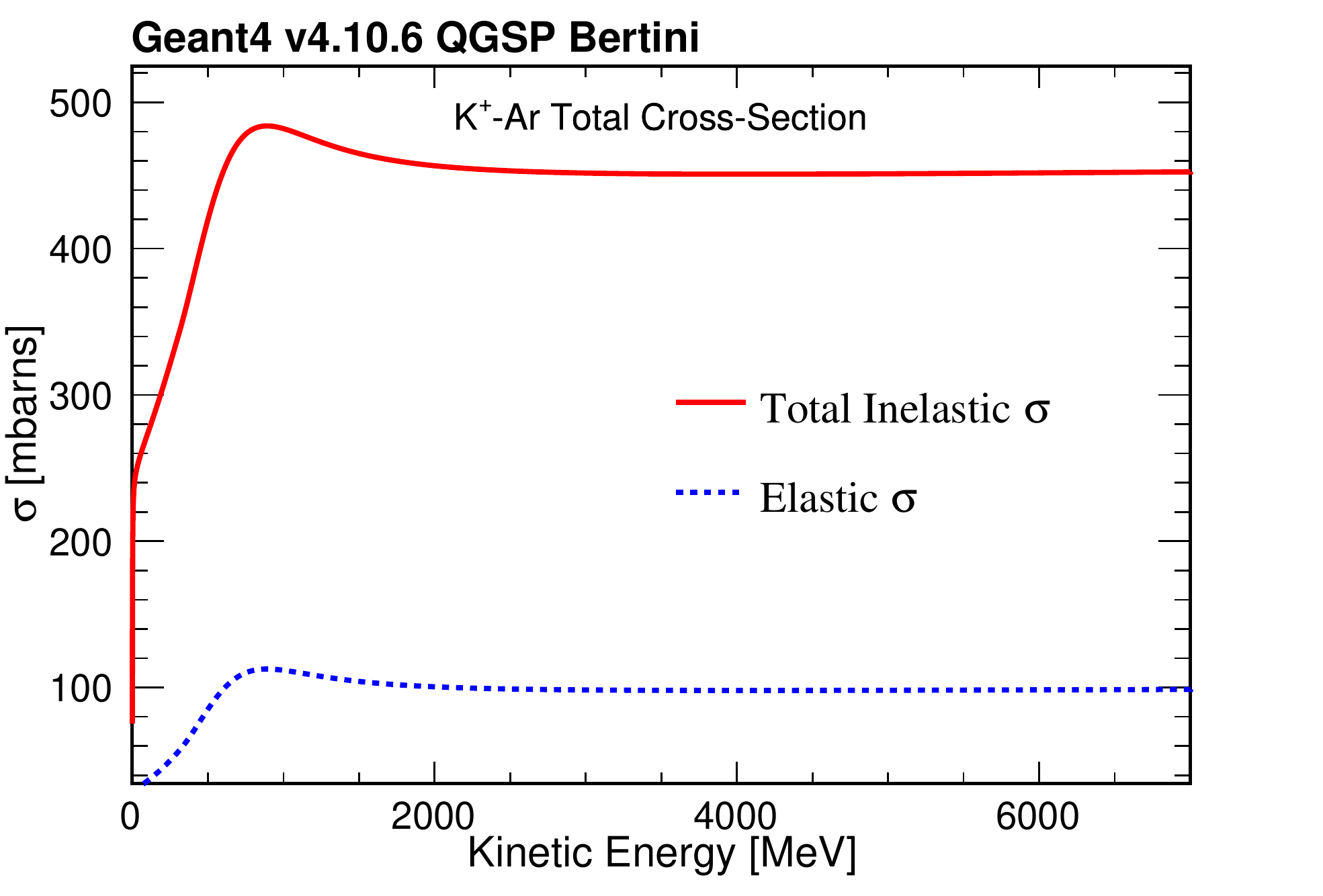}
    \caption{\textsc{Geant4} predicted total inelastic cross section and elastic cross section of positively-charged kaons on argon as a function of kinetic energy \cite{GEANT4, geant4IEEE, ALLISON2016186}. Predictions made using interfaces in Reference~\cite{Calcutt_2021}.}
    \label{fig:kaonPred}
\end{figure}

In this work, the kaon-argon total inelastic cross section is reported as a function of kaon energy within the limits of the detection threshold, described in Section~\ref{sec:unfold}. Figure~\ref{fig:kaonPred} shows the total inelastic and the elastic cross section predicted by the \textsc{Geant4} Bertini cascade model~\cite{GEANT4, geant4IEEE, ALLISON2016186}. Charged kaons produced by the beam with kinetic energies of approximately 4.5 to 7 GeV are capable of reaching the liquid argon of ProtoDUNE-SP. Using the \textsc{Geant4} prediction from Figure~\ref{fig:kaonPred}, the simulated total inelastic cross section at the relevant energies should be approximately 450 millibarns (mbarns).  

 Section~\ref{sec:pdune} discusses ProtoDUNE-SP more broadly, and Section~\ref{sec:reco} outlines the simulation and reconstruction of ProtoDUNE-SP data. Section~\ref{sec:unfold} explains the \textit{thin slice method} used in this measurement. This method divides the detector into thin targets, referred to as \textit{thin slices}, using the wires of the LArTPC to demarcate the slices. An \textit{incident} slice is counted if a particle reaches a particular wire. Within that slice, it may also interact on the argon, which means the slice contains both an incident and an \textit{interacting} slice. After an interacting slice is detected, the counting for the event stops as the outgoing particles have unknown identities and energies. The cross section is measured using the counts of the incident and interacting slices as a function of kinetic energy.

Section~\ref{sec:kaonSel} describes the selection of candidate kaon interaction events, and Section~\ref{sec:calo} shows energy-related measurements using selected kaons. Section~\ref{sec:results} reports the kaon-argon cross section with comparisons to models. Section~\ref{sec:uncertainties} discusses the evaluations of the statistical and systematic uncertainties.

\section{ProtoDUNE-SP and the H4-VLE Beamline}
\label{sec:pdune}

ProtoDUNE-SP is a 770-ton liquid argon detector with two TPCs, each with a drift distance of 3.6 m~\cite{Abed_Abud_2022}. The detector contains six readout wire planes called Anode Plane Assemblies (APAs). These APAs contain three readout wire planes, the U, V, and X wire planes, and are 6.2 m high, 2.3 m wide, and 0.1 m thick~\cite{Abed_Abud_2022}. The U and V wires are the first two planes and detect drifting electrons via the currents induced on the wires as the charges drift past them, creating bipolar signals. The X wires, known as collection wires, have unipolar signals where the drifting electrons collect on the wires and stop drifting in the TPC~\cite{Abed_Abud_2022}.

The U, V, and X wires are oriented 35.7$^{\circ}$, -35.7$^{\circ}$, and 0$^{\circ}$ relative to the vertical direction, respectively. The pitch between wires is 0.467 cm for induction wires and 0.479 cm for collection wires. Each APA has 960 X wires, 800 U wires, and 800 V wires. Three APAs lie on one wall of the cryostat, and the other three APAs lie on the opposite wall of the cryostat. These APAs are 7.2 m away from each other, and the Cathode Plane Assembly (CPA) sits midway between the two separate walls of APAs. The CPA provides a high voltage of 180 kV, leading to a nominal electric field strength of 500 V/cm across the 3.6 m separating each APA from the CPA, which allows the ionization electrons to drift to the APAs. The H4-VLE beam pipe connects to the upstream face of LArTPC via a low-density beam plug that allows the beam to enter without scattering off the material in the cryostat~\cite{abi2020first, Abed_Abud_2022}. 

The beam only enters one TPC of the detector. The beam side of the detector contains APAs separated by electron diverters intended to improve the charge-collection efficiency for electrons drifting near the gap between neighboring APAs. Unfortunately, these electron diverters exhibited high-voltage shorts and were left electrically grounded during operations, distorting the track images and causing some loss of collected charge. 

As a surface-based detector, ProtoDUNE-SP is exposed to an intense flux of cosmic-ray muons, which create electron-ion pairs in the detector. The argon ions drift slower than the ionization electrons, leading to an excess of ions around the surface of the detector. The excess of ions creates a space charge effect that alters the local electric field, leading to distorted calorimetry and tracking~\cite{Palestini}. 

A calibration of the space charge effect is completed by measuring the tracking distortions on the surfaces of the detector, where the effect is maximal, with cosmic-ray muon data~\cite{abi2020first}. The distortions measured are then used to correct for local electric field fluctuations by using a linearly interpolated three-dimensional map. An ``inverted'' map using these data measurements is used to recreate the space charge effect in simulation. The original three-dimensional map is utilized to calibrate this effect in simulation.

From September 2018 to early November 2018, the H4-VLE  beamline settings were adjusted to emit positively-charged particles at 0.3, 0.5, 1, 2, 3, 6, and 7 GeV/$c$ beamline momentum settings. The beamline trigger operates at a rate of 25 Hz, which qualitatively translates to beam particles being observed one-at-a-time within ProtoDUNE-SP. The beam consists of positively charged protons, positrons, kaons, pions, and muons. The beam particle species is identified using a time-of-flight system and Cherenkov detectors. The beam particle momentum is measured from the bend of the particle's trajectory through a well-known magnetic field using data from tracking fibers~\cite{abi2020first,protoDUNEBeamPerf}. At 6 GeV/$c$ and 7 GeV/$c$ beam momentum settings, particle identification comes from only the Cherenkov detectors, explicitly requiring a signal in the high-pressure Cherenkov detector but no signal in the low-pressure Cherenkov detector~\cite{abi2020first}. 

\section{Simulation and Reconstruction}
\label{sec:reco}

A simulation of the beam, including its transport to and through the LArTPC, is implemented using \textsc{Geant4}~\cite{GEANT4, geant4IEEE, ALLISON2016186}, with the entire CERN H4-VLE facility simulated from the primary beam to the tertiary beam that reaches ProtoDUNE-SP~\cite{protoDUNEBeamPerf}. The selection of kaon inelastic scattering events starts with the beamline instrumentation discussed in Section~\ref{sec:pdune}. An event is defined as any time the beamline instrumentation has a signal recorded by the high-pressure Cherenkov detector and the absence of a signal recorded by the low-pressure Cherenkov detector~\cite{abi2020first}. 

The rest of the selection steps rely on information from the reconstruction of tracks and showers in the TPC to select relevant events. Additionally, the beamline instrumentation also has tracking fibers to reconstruct a beam track that can be extrapolated to the TPC~\cite{abi2020first,protoDUNEBeamPerf}.

ProtoDUNE-SP uses the Pandora multi-algorithm reconstruction package to identify the beam particle, reconstructing particle hierarchies using pattern recognition~\cite{abi2020first,Marshall_2015,pandoraProtoDUNE}. It then employs a boosted decision tree to select beam particle candidates that enter through the beam pipe and beam plug into the liquid argon detector. A full description of the software used in ProtoDUNE-SP is given in References~\cite{abi2020first, pandoraProtoDUNE}. 

\begin{figure}
    \centering \includegraphics[width=0.45\textwidth]{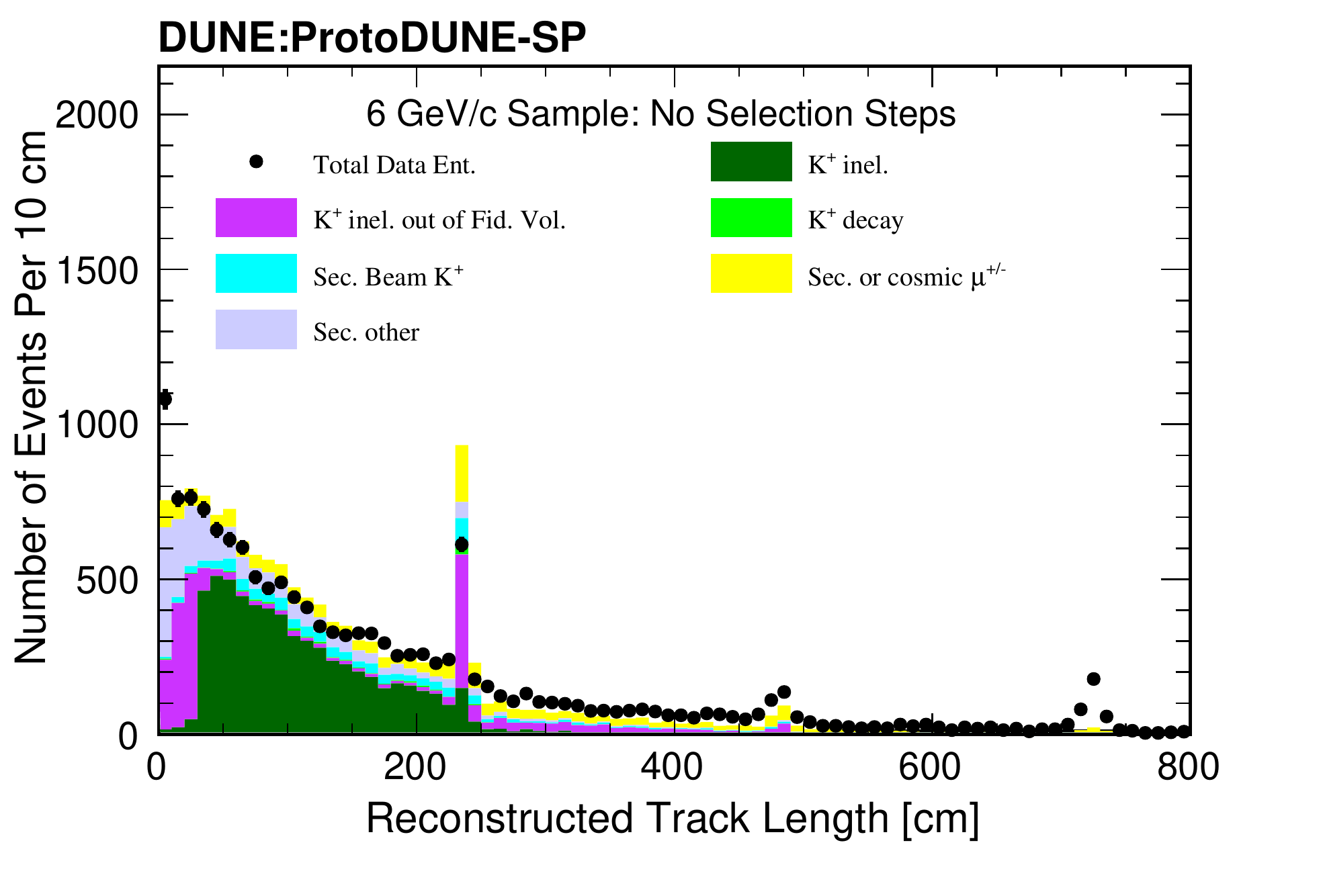}
    \caption{Reconstructed track length for simulation and data without any TPC-related selection steps for data and simulation at the 6 GeV/$c$ beamline setting. Only statistical uncertainties from the data are shown. The statistics of the simulation are scaled to match the normalization of all data events, including those without a reconstructed track in the TPC.}
    \label{fig:kaonNoSelTrkLen}
\end{figure}

Figure~\ref{fig:kaonNoSelTrkLen} shows the observed and simulated distributions of the reconstructed track lengths for events with a beam kaon, as reported by only the beamline instrumentation, for the 6 GeV/$c$ samples. The corresponding distributions for the reconstructed track lengths and all other event selection distributions for the 7 GeV/$c$ samples showed similar agreement and are included in Appendix~\ref{sec:appendix}. The spikes in Figure~\ref{fig:kaonNoSelTrkLen} at around 230 cm and 460 cm correspond to the electron diverters that inadvertently break tracks between APAs in the detector, as discussed in the previous section, with the last spike at around 700 cm corresponding to the end of the active volume. Most TPC tracks are secondary particles without any TPC-related selection steps. An excess of short reconstructed track lengths is observed in the data, likely driven by background secondary particles.

The interaction point, or track endpoint, is determined using clustering and vertex-finding algorithms that are almost identical to those from the MicroBooNE reconstruction and are described in detail in Reference~\cite{Pandora_UBooNE}. Initial clusters are created in a way to ensure that the clusters only contain signals from one particle but are unlikely to contain all TPC signals from that particle. Numerous algorithms are applied to topologically merge clusters with the goal of one cluster containing all of the activity from one particle. In addition, algorithms are applied to split clusters if kinks are found or where the topology suggests that there may be contributions from multiple particles. These clusters are classified as either tracks or showers based on their topologies. Candidate 3D interaction points are produced by comparing pairs of clusters from two 2D views and reconstructing their start and end points as candidate interaction points. A boosted decision tree is then used to select the candidate vertex most likely to be the point where the beam particle interacts. Using Pandora vertex-finding algorithms, interaction points are observed. The signal process is an inelastic interaction of the incident kaon. An inelastic interaction is defined as any process where either:

\begin{itemize}
    \item the angle between the beam kaon and leading outgoing particle is greater than 11 degrees
    \item two or more particles emerge from the interaction point.
\end{itemize}

The kinetic energy threshold for observing a final-state proton or charged kaon in the detector is 40 MeV, and for a charged pion, it is 20 MeV. We apply these restrictions to our signal definition.

\section{Methodology}
\label{sec:unfold}

The cross section measurements presented in this paper use the \textit{thin slice method} pioneered by the LArIAT experiment~\cite{gramellini2018measurement, elena}. The approach treats the detector as a series of thin argon targets (slices). The number of surviving particles ($N_{\rm{surv}}$) is:

\begin{equation}
N_{\rm{surv}}\left(d\right)=N_{\rm{inc}}\exp\left(-d/l\right)=N_{\rm{inc}}\exp\left(-\sigma  d n \right)
\label{eqn:surv}
\end{equation}
where $N_{\rm{inc}}$ is the number of incident particles, $d$ is the distance traveled, and $l=(n\sigma)^{-1}$ is the interaction length of a kaon in argon, where $n$ is the number density and $\sigma$ is the cross section~\cite{gramellini2018measurement, elena}. 

 A natural way to develop slices in a LArTPC with a wire readout is to use the individual wires to demarcate thin target slices from one another. Therefore, a slice is a three-dimensional box of argon between wires. For each particle, the incident energy at each thin slice is estimated. The total number of particles at each incident energy ($N_{\rm{inc}}$) is counted, as are the total number of interactions ($N_{\rm{int}}$). Regardless of whether or not there was an interaction, if an energy deposit from the kaon is registered in a thin target slice, then the incident energy of the kaon is counted ($N_{\rm{inc}}$). The cross section, using Equation~\ref{eqn:surv}, is:

\begin{equation}
    \sigma\left(E_{\rm{kin}}\right)=\frac{M_{\rm{Ar}}}{N_{\rm{A}} r \rho} \ln\left[\frac{N_{\rm{inc}}(E_{\rm{kin}})}{N_{\rm{inc}}(E_{\rm{kin}})-N_{\rm{int}}(E_{\rm{kin}})}\right]
    \label{eqn:xsecthin}
\end{equation}
where $E_{\rm{kin}}$ is the kinetic energy of the particle, $N_{\rm{A}}$ is the Avogadro constant, $M_{\rm{Ar}}$ is the atomic mass of argon, $\rm{\rho}$ is the density of liquid argon, and $r$ is the three-dimensional distance the particle travels from one wire to the next. This value is 0.498 cm, given the wire spacing between the collection plane wires of 0.479 cm and that the beam travels at a 16-degree angle in the detector~\cite{Diurba_thesis,gramellini2018measurement, elena}.

The kinetic energy at a given slice ($E_{\rm{kin,j}}$) is defined as: 

\begin{equation}
    E_{\rm{kin,j}}=E_{\rm{kin,beam}}-\sum^{j-1}_{i=0} \Delta E_i
       \label{eqn:eslice}
\end{equation}
where $E_{\rm{kin,beam}}$ is the initial beam particle kinetic energy and $\Delta E_i$ is the energy lost in slice $i$. The total $\Delta E$ is summed from all slices up to slice $j$.

Background subtractions, unsmearing, and efficiency corrections are required to convert the measured interaction and incident spectra into a cross section. These corrections are applied via RooUnfold with unfolding done using a Bayesian-like unfolding algorithm implemented based on Richardson-Lucy deconvolution~\cite{DAGOSTINI1995487,Richardson,Lucy,RooUnfoldPaper,roounfoldRepo}. The background subtractions, unsmearing, and efficiency corrections are done by RooUnfold. These corrections are applied on the incident and interacting spectra separately, an approach similar to that previously used by LArIAT~\cite{elena}. These unfolded distributions of the incident and interacting slices are then used in Equation~\ref{eqn:xsecthin} to measure the cross section as a function of kinetic energy.

\section{Event Selection}
\label{sec:kaonSel}

There are three event selection steps to select candidate kaons and an additional step to select a candidate kaon with an inelastic interaction. They include the following selection steps for events where the beamline trigger reports a kaon candidate:

\begin{itemize}
    \item The event must have a reconstructed TPC track.
    \item The endpoint of the TPC track must enter the fiducial volume by being at least 30 cm downstream of the start of the active volume of the detector. This selection step is motivated by significant inefficiencies and impurities in correctly identifying and reconstructing the beam particle with a TPC track in the first 30 cm of the detector. 
    
    \item The TPC track must be matched to the trajectory of the beam track from the beamline instrumentation. A match requires that their positions and angles agree within three times the resolutions of these measurements at the start of the fiducial volume.

\end{itemize}

\begin{figure}
    \centering
     \includegraphics[width=0.42\textwidth]{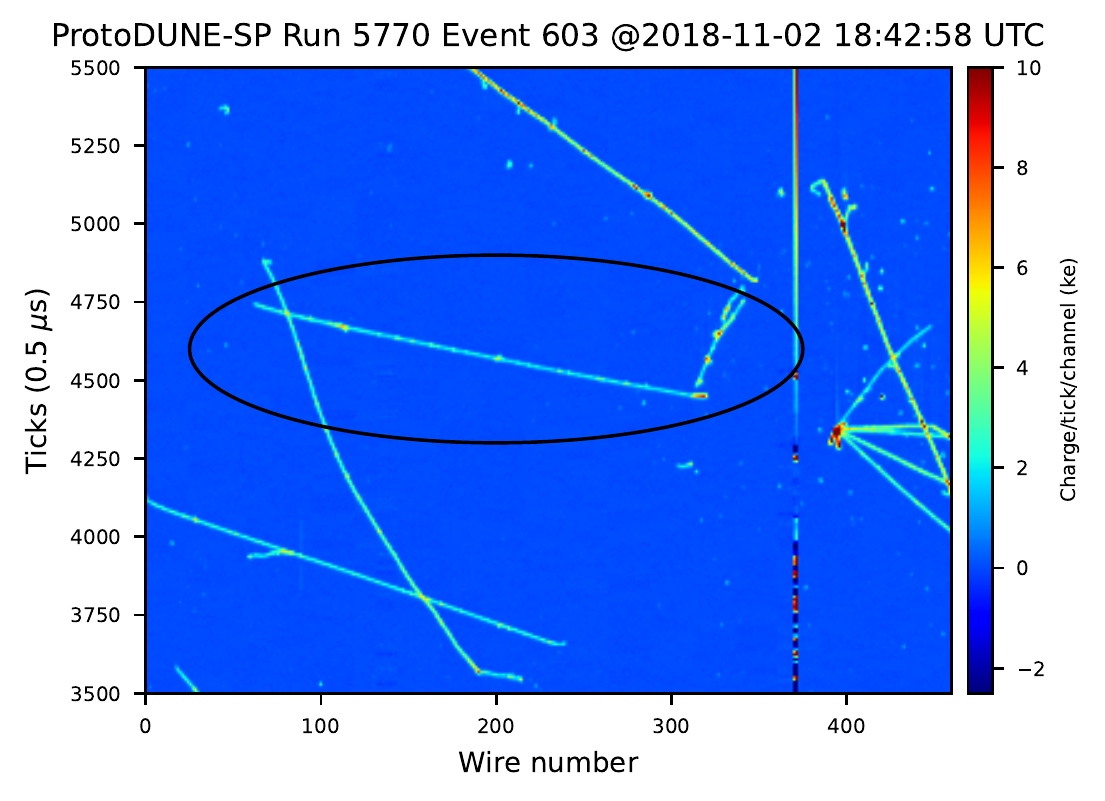}   \includegraphics[width=0.42\textwidth]{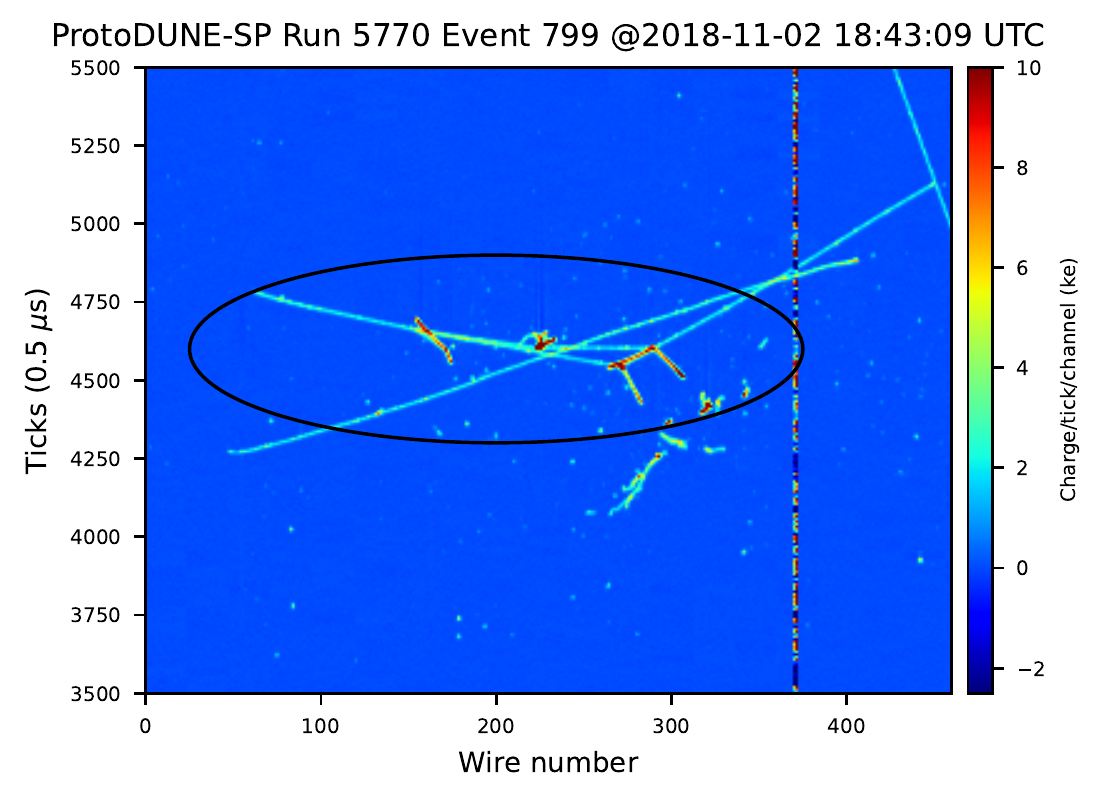}
     \includegraphics[width=0.42\textwidth]{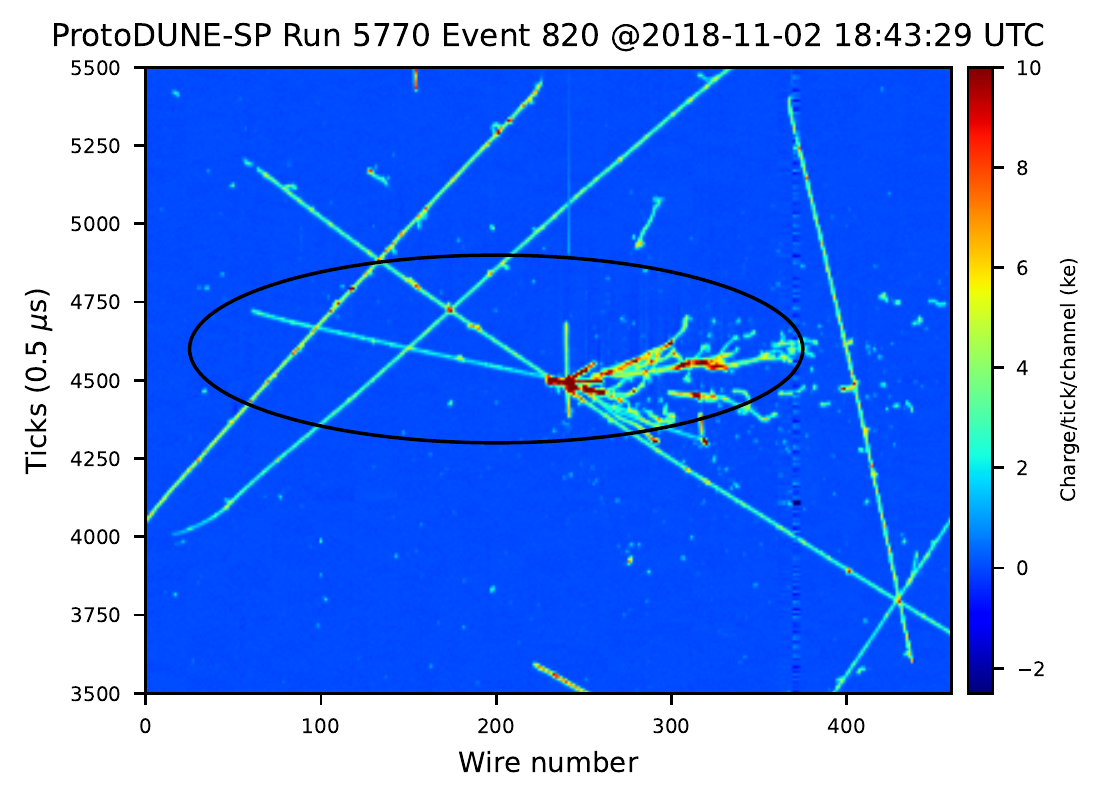}
    
    \caption{Three candidate event displays of selected beam kaons that may inelastically interact on the argon using data from early November 2018. The beam travels from the left to the right at an angle of approximately 16 degrees. Cosmic-ray muons can be seen in the foreground and background of the beam event, and a non-functioning wire can be observed near wire 370.}
    \label{fig:eventDisplays}
\end{figure}

Because the electron diverters tend to break tracks, as discussed in Section~\ref{sec:pdune}, only the interaction and incident slices contained before the point of 220 cm across the detector length are considered in the cross-section measurement. This is the final step. At each collection wire, the kaon energy is estimated per Equation~\ref{eqn:eslice}, and the kaon either undergoes an interaction or does not. Thus, for each incident particle, we observe many ``slices'' and record the interaction as a function of energy. The interaction point, or vertex, identification occurs through Pandora as described in Section~\ref{sec:reco}. Event displays of some selected kaon inelastic interaction candidates are shown in Figure~\ref{fig:eventDisplays}. In these events, the beam enters the TPC at time tick 4750, where a time tick represents the 500 ns sampling intervals of the analog-to-digital converters for the wires, and then travels over 50 cm before interacting with the argon. The beam particles, highlighted by the black ovals, travel in approximately straight lines from the left to the right before scattering, creating complicated final states with many showers. The top two event displays show little shower activity, indicating they may be $1K^{+}+X$ candidate events. The third event display shows a complex interaction with many showers and tracks in the final state. 

\begin{figure}
    \centering 
        \includegraphics[width=0.45\textwidth]{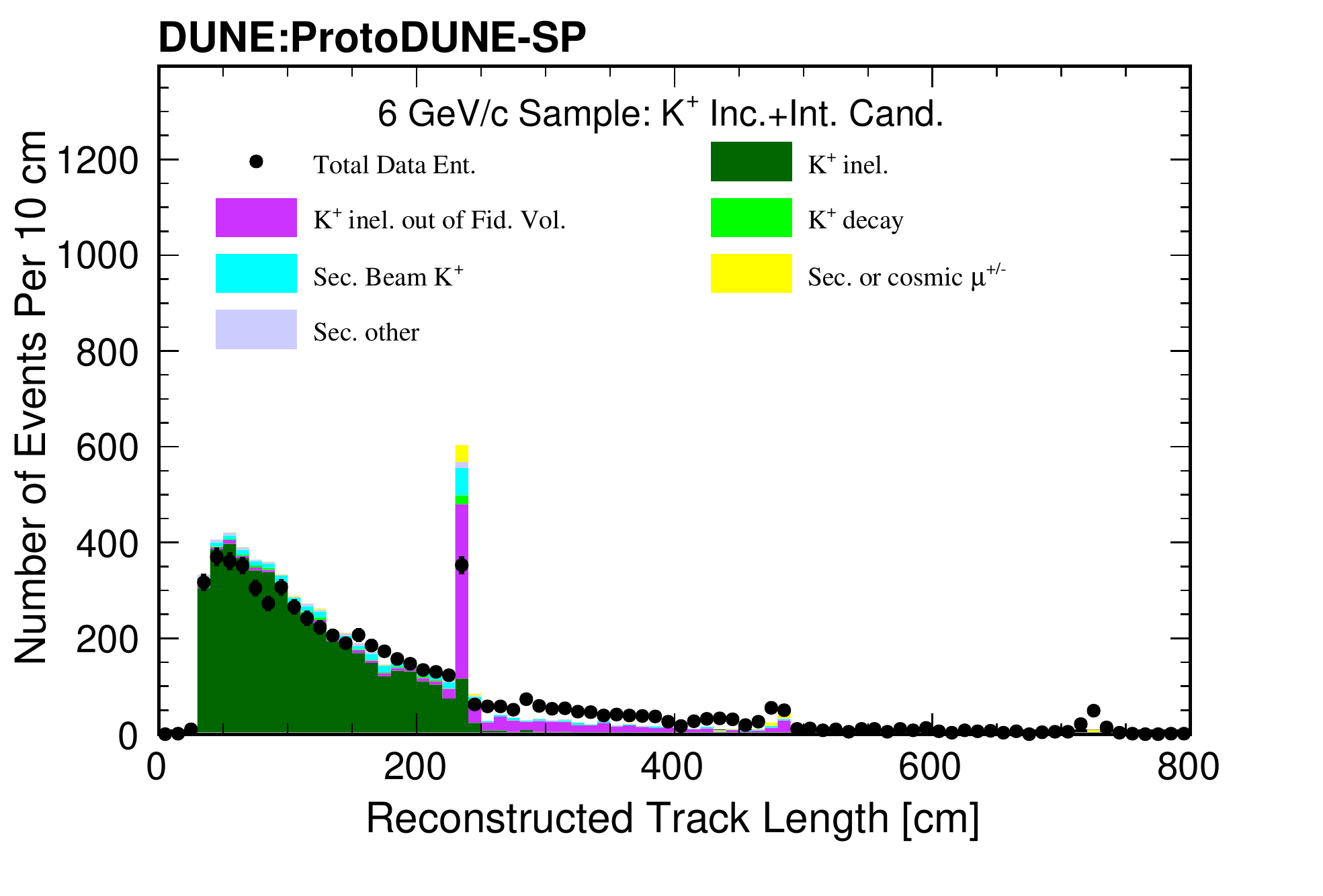}
        \includegraphics[width=0.45\textwidth]{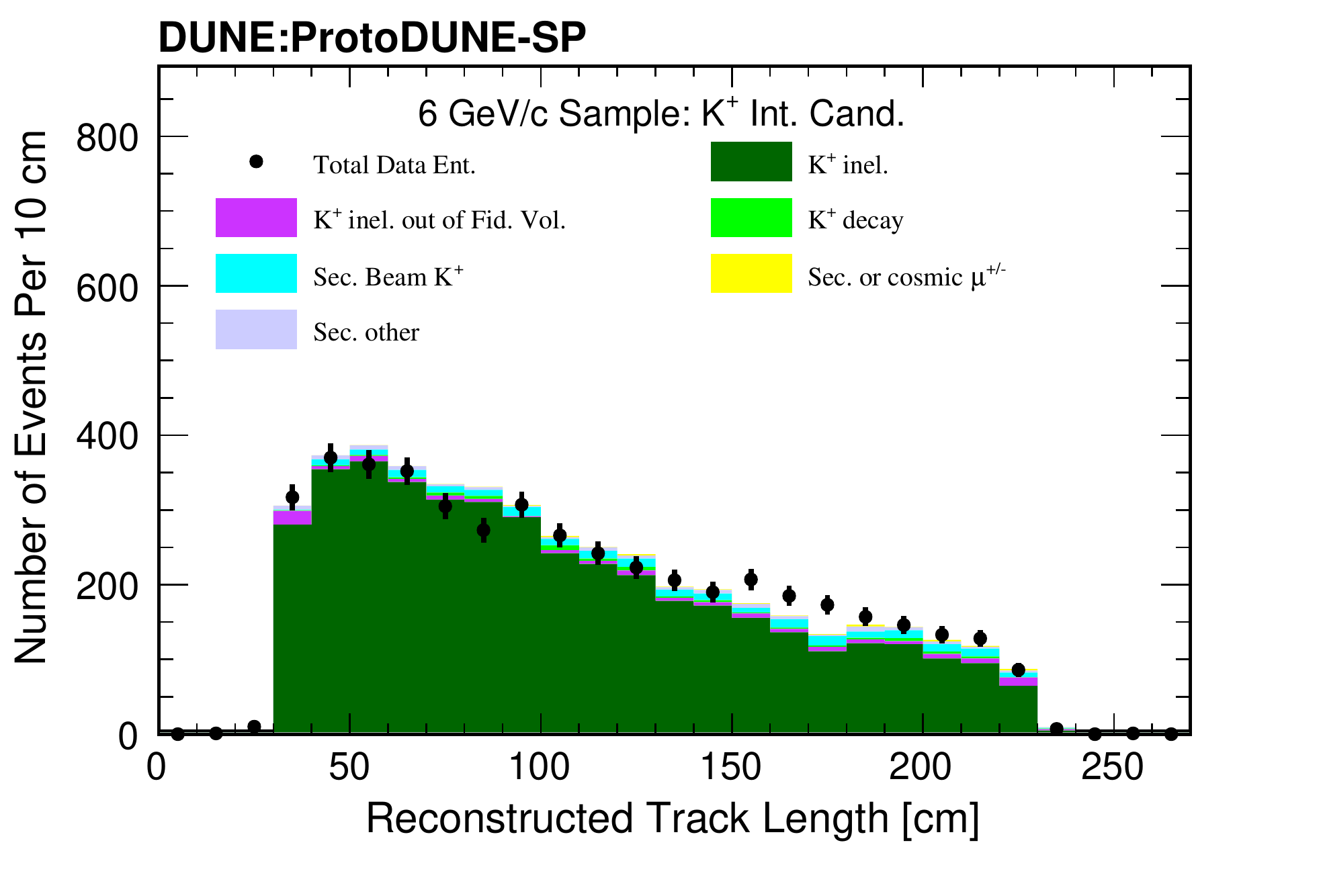}
    \caption{Reconstructed track length for simulation and data of the 6 GeV/$c$ beamline setting for selected kaons (top) and for selected kaons that interact within the fiducial volume (bottom). Only statistical uncertainties are shown for the data, and the statistics from the simulation are scaled to match those from the data.}
    \label{fig:kaonSelAllTrkLen}
\end{figure}

Figure~\ref{fig:kaonSelAllTrkLen} shows the distributions of reconstructed track lengths for selected TPC tracks that will form the incident and interacting slice spectra from the 6 GeV/$c$ beamline setting. Secondary kaons, which are byproducts of true beam kaons interacting off the argon and travel with some unknown kinetic energies, are the most significant background for the event selection. As these secondary kaons will have similar characteristics to beam kaons, they are an irreducible background. The breakdown of the data and simulation samples through each selection step are shown in Table~\ref{tab:standardStats}.

\begin{table*}[htb]

    \caption{Information on the fractions of the samples remaining for data and simulation after each selection step from the left (beamline reports a candidate kaon) to the right (candidate kaon has an interaction in the fiducial volume).} 
    \centering

    \begin{tabular}{|c|c|c|c|c|c|} \hline 
   Selection step & Beam & TPC track & Fiducial & Beam-TPC match & Contained  \\ \hline 
6 GeV/$c$ data & 100.0\% & 58.0\% & 46.0\% & 25.4\% & 18.6\% \\ \hline 
7 GeV/$c$ data & 100.0\% & 55.6\% & 44.8\% & 27.3\%  & 19.5\% \\ \hline 
6 GeV/$c$ sim total  & 100.0\% & 55.0\% & 44.7\% & 29.1\% & 23.2\% \\ \hline 
6 GeV/$c$ sim signal  & 24.9\% & 24.4\% & 24.0\% & 21.8\% & 20.9\% 
\\ \hline
6 GeV/$c$ sim bkg  & 75.1\% & 30.6\% & 20.7\% & 7.3\% & 2.2\% \\ \hline
7 GeV/$c$ sim total & 100.0\% & 45.1\% & 36.5\% & 24.0\% & 19.1\% \\ \hline 
7 GeV/$c$ sim signal  & 20.9\% & 20.4\% & 20.0\% & 18.3\% & 17.5\% \\ \hline 
7 GeV/$c$ sim bkg & 79.1\% & 24.7\% & 16.5\% & 5.6\% & 1.5\% \\ \hline

    \end{tabular}
    \label{tab:standardStats}
\end{table*}

\begin{figure}
    \centering
    \includegraphics[width=0.45\textwidth]{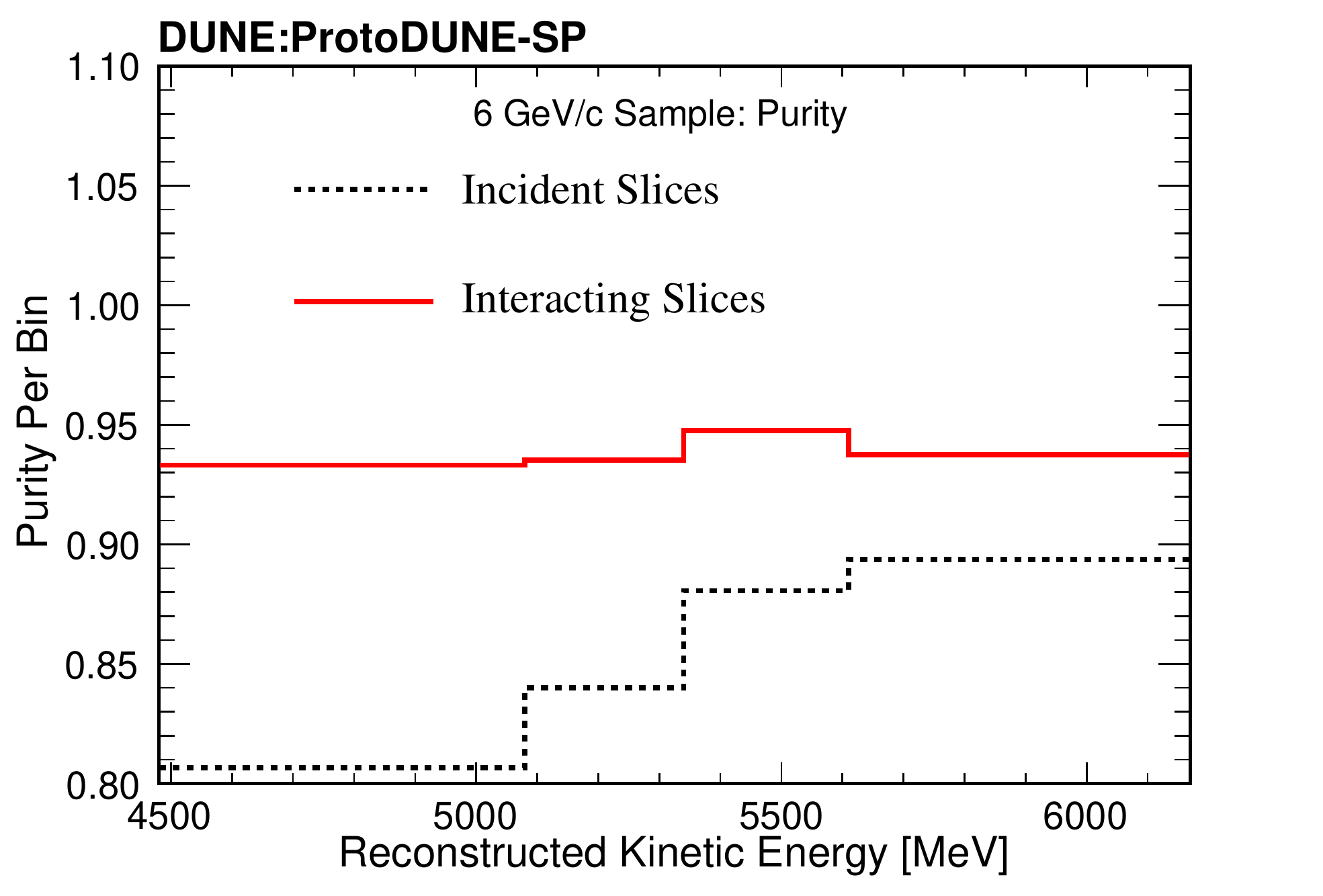}
    \includegraphics[width=0.45\textwidth]{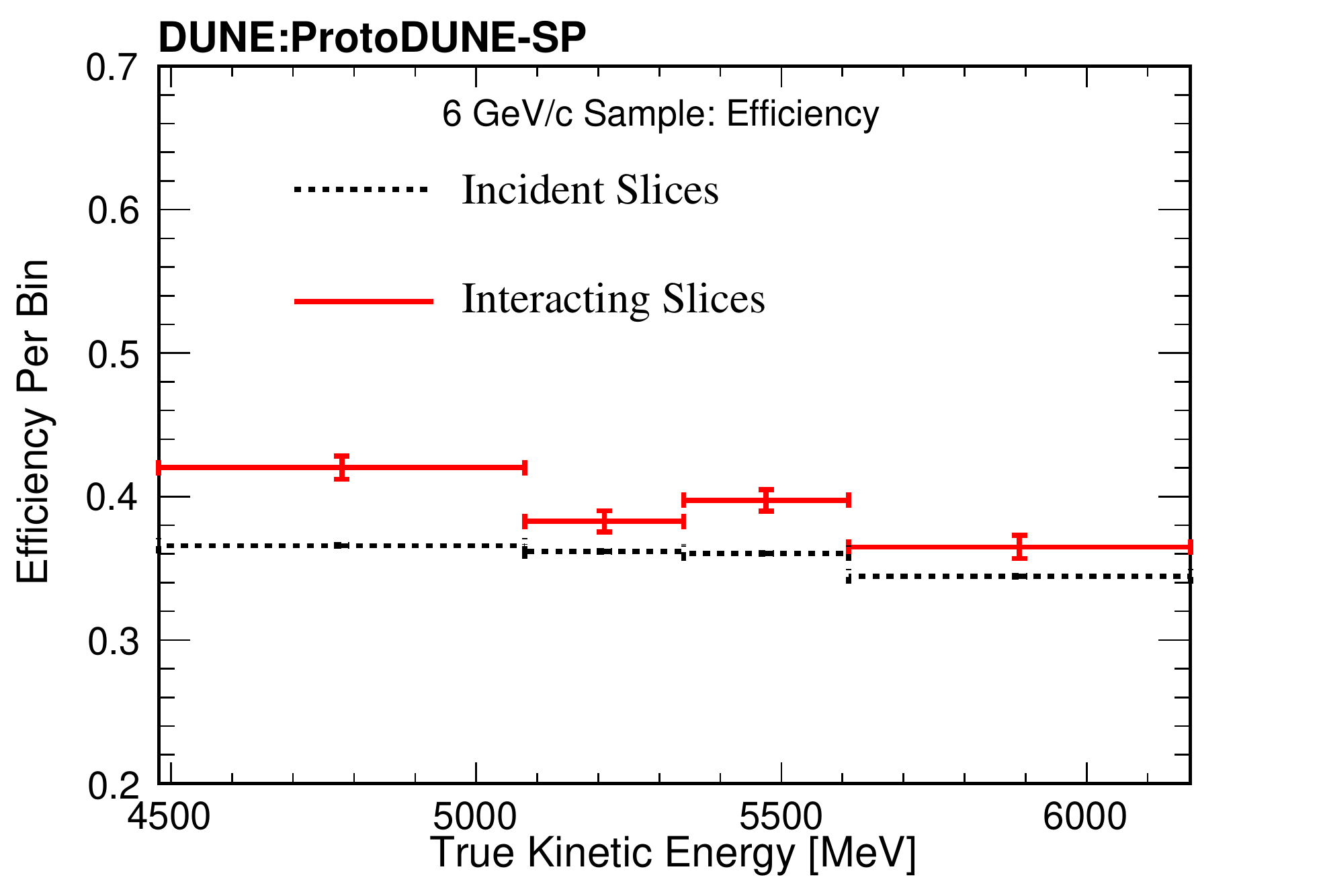}
    \caption{Purity (top) and efficiency (bottom) for each bin using the event selection for the 6 GeV/$c$ simulation sample.}
    \label{fig:purEff}
\end{figure}

The selection efficiency and purity are evaluated as a function of kinetic energy from simulation. An inefficiency in measuring a slice of kinetic energy occurs when no TPC track corresponds to the beam particle in the slice. A background slice occurs when there is a TPC track in a slice that the true kaon does not reach. The definition for a background slice is used regardless of whether the TPC track is from a true kaon or not, which allows unfolding to fully recover the truth distribution after unfolding. Results are shown in Figure~\ref{fig:purEff}. The purity is close to 95\% for interacting slices and 85\% for incident slices. The lower purity is because a single background particle entering the detector contributes to many non-interacting slices, but only the first interaction is recorded. The efficiency varies between 35-40\% as a function of energy. The inefficiencies are dominated by events with a true kaon in the fiducial volume, but the event did not have a TPC track identified as the beam particle.

\section{Energy Measurements and Binning} \label{sec:calo}

\begin{figure}
    \centering
    \includegraphics[width=0.45\textwidth]{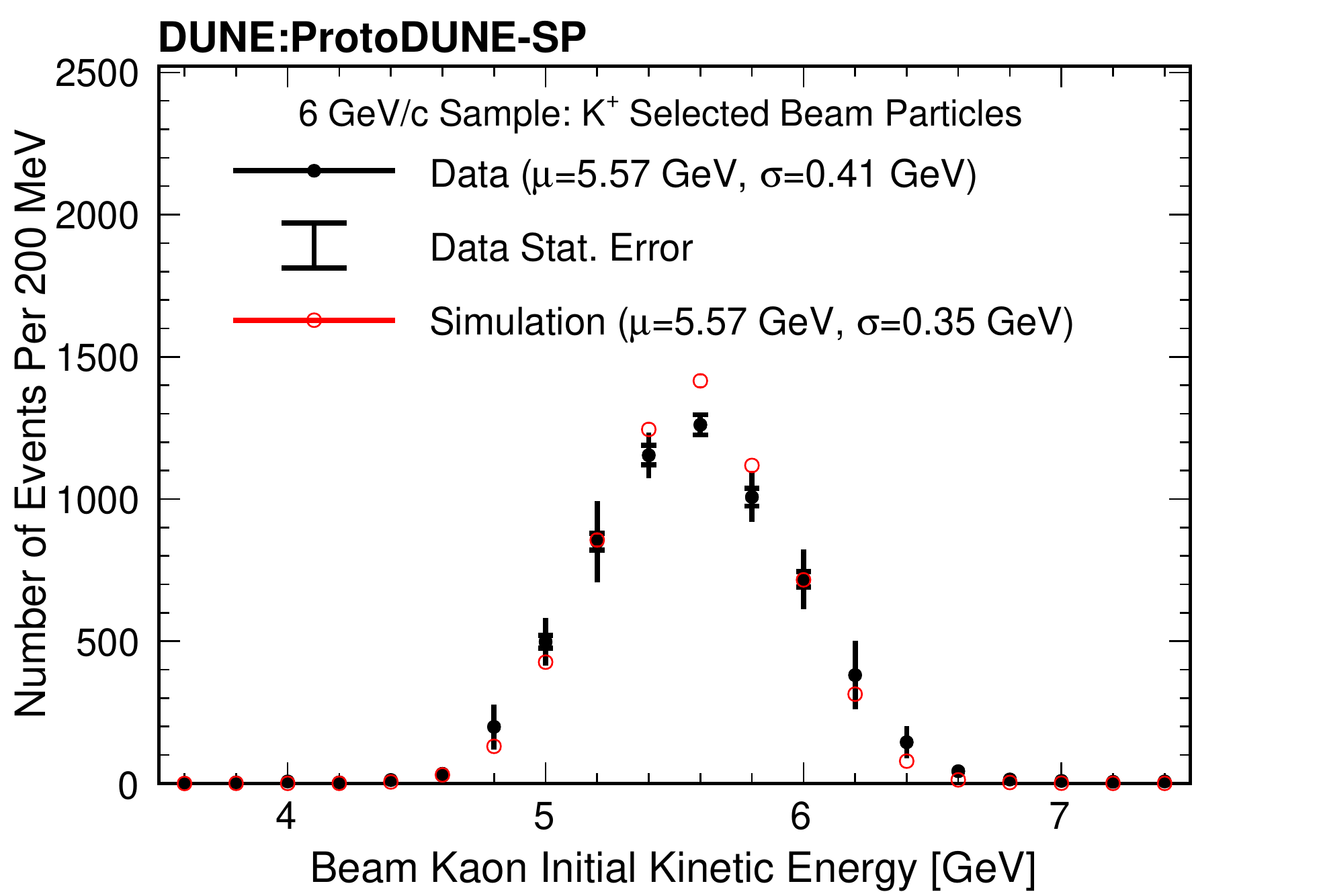}

    \caption{Initial beam particle kinetic energy as measured by the beamline instrumentation for selected kaon candidate tracks for the 6 GeV/$c$ beamline setting. Both systematic and statistical uncertainties are shown.}
    \label{fig:beam}
\end{figure}

As referenced in Equation~\ref{eqn:eslice}, the initial kinetic energy is determined using measurements from the beamline instrumentation. Figure~\ref{fig:beam} displays the beamline kinetic energy measurements of the selected beam kaons. The systematic uncertainty error bars are found by shifting the data distribution by the 1.2\% kinetic energy modeling uncertainty of the beamline simulation, which will be discussed in greater detail in Section~\ref{sec:uncertainties}.

\begin{figure}
    \centering
            \includegraphics[width=0.45\textwidth]{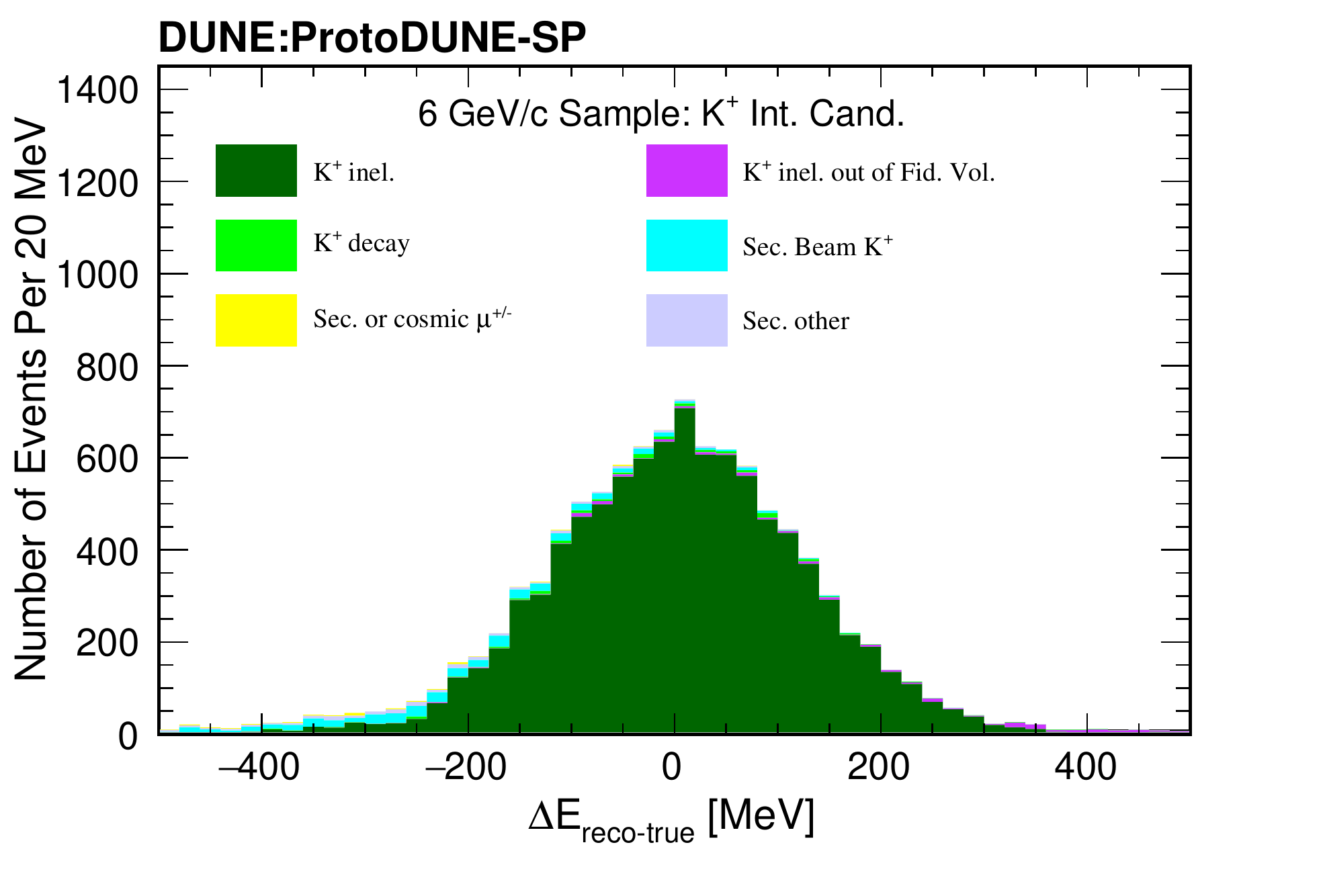} 
    \caption{Kinetic energy resolution at the interaction point of beam particles that pass all selection criteria for interacting kaons in the 6 GeV/$c$ simulation sample. The distribution has a mean energy bias of 0.50 MeV. The distribution is not scaled to the statistics in the data.}
    \label{fig:res}
\end{figure}

The TPC calorimetry is calibrated by applying corrections to the electric field variations, corrections for the spatial variations, and an overall charge scale using through-going and stopping cosmic-ray muons~\cite{abi2020first}. The energy resolution was evaluated to find the binning resolution, and done by measuring the difference between the true and reconstructed kinetic energies at the interaction points in the simulation. The resolution is measured to be 124 MeV, as seen in Figure~\ref{fig:res}.

\begin{figure}
    \centering
    \includegraphics[width=0.4\textwidth]{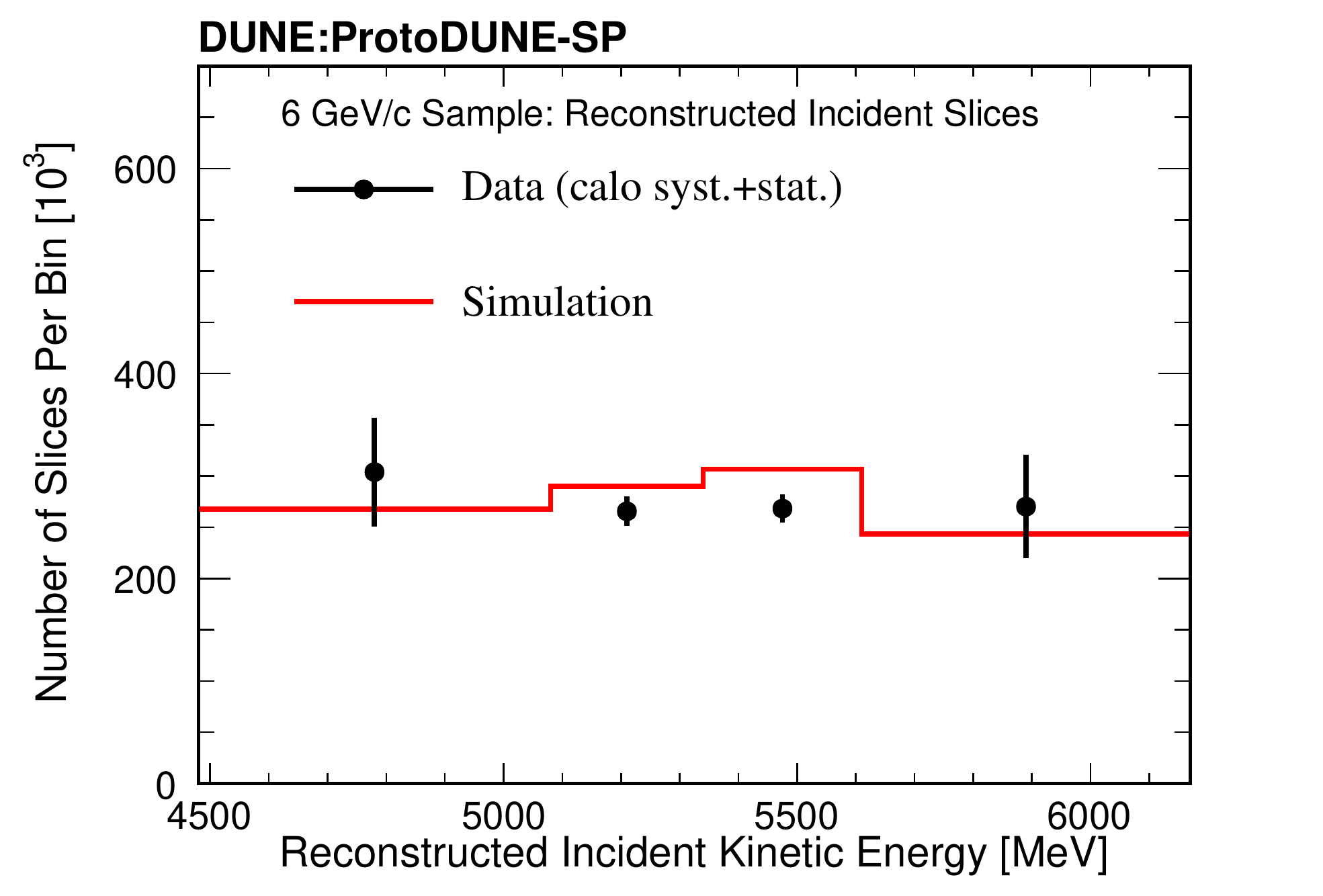}
        \includegraphics[width=0.4\textwidth]{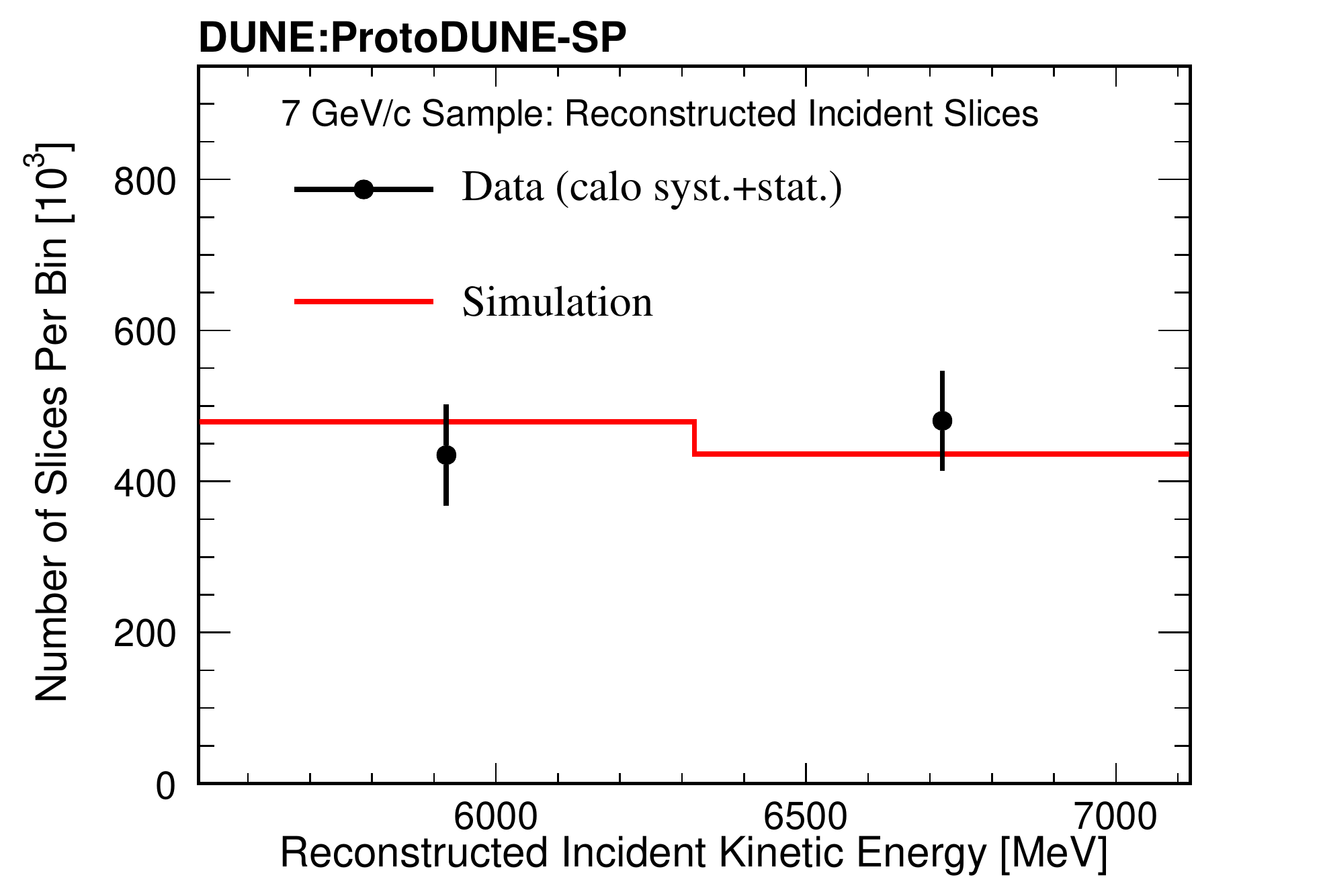}
    \caption{Reconstructed incident slice distributions between the data and simulation for the 6 GeV/$c$ beamline setting (top) and the 7 GeV/$c$ beamline setting (bottom). A calorimetric slice-by-slice uncertainty of 3\% and a beam kinetic energy scale uncertainty of 1.2\% are applied to the data. Statistics for the simulation are scaled to match the normalization from the data.}
    \label{fig:recoInc}
\end{figure}

\begin{figure}
    \centering
    \includegraphics[width=0.4\textwidth]{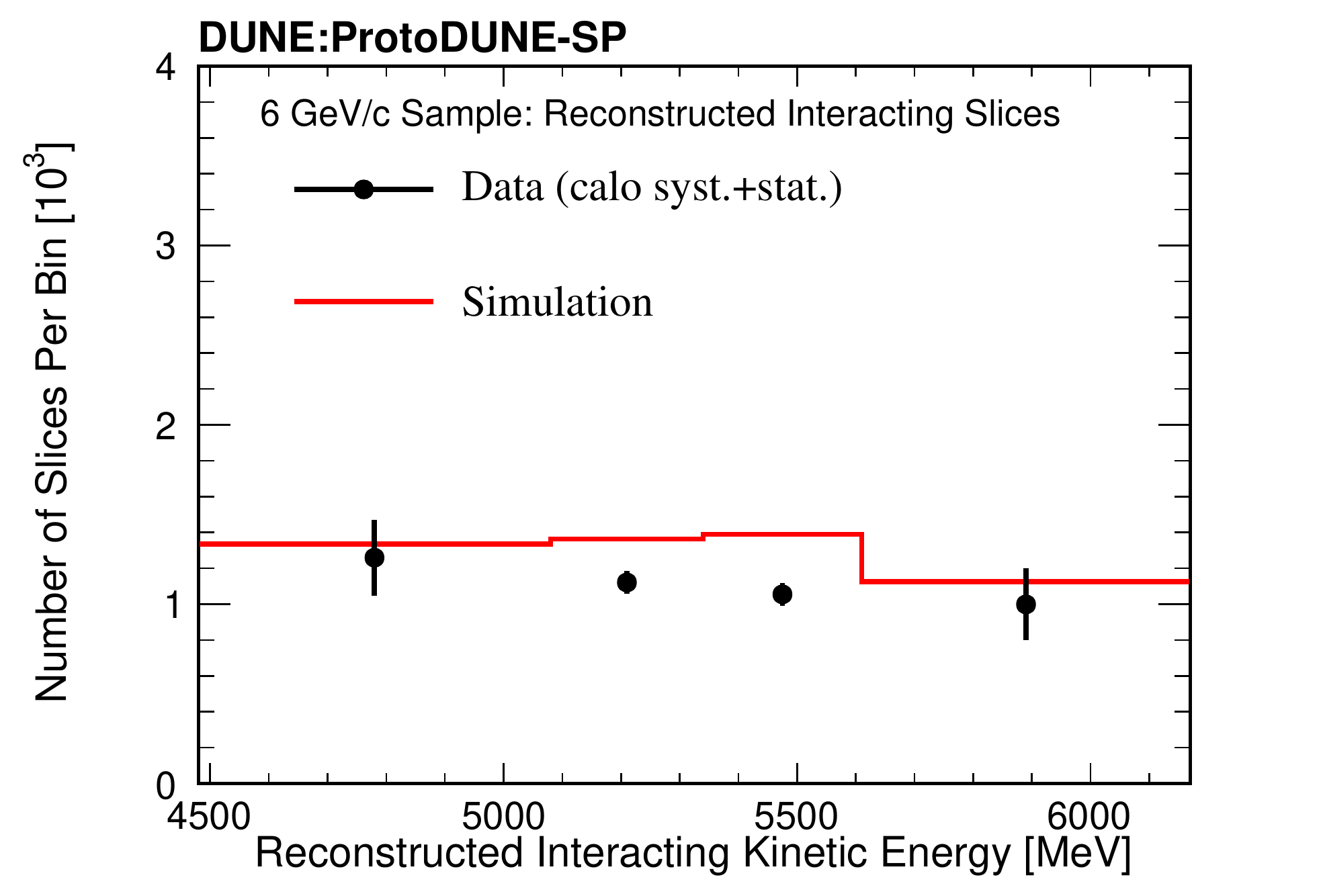}
        \includegraphics[width=0.4\textwidth]{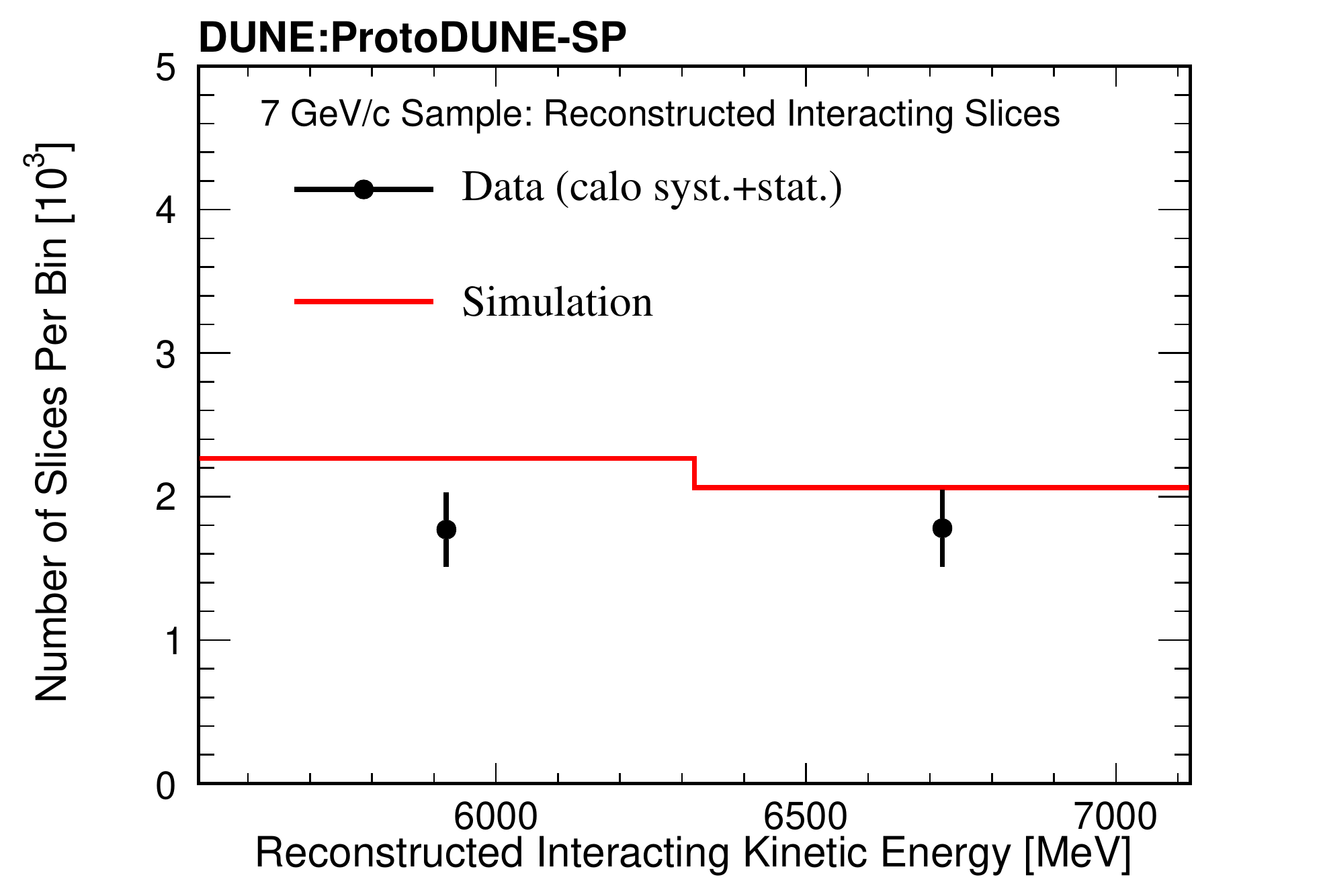}
    \caption{Reconstructed interacting slice distributions between the data and simulation for the 6 GeV/$c$ sample (top) and the 7 GeV/$c$ sample (bottom). A calorimetric slice-by-slice uncertainty of 3\% and a beam kinetic energy scale uncertainty of 1.2\% are applied to the data. Statistics for the simulation are scaled to match the normalization of incident slices from the data.}
    \label{fig:recoInt}
\end{figure}

The binning of the analysis was developed to address the kinetic energy resolution and ensure equal statistics in each bin for the reconstructed interacting slice distributions in the data sample for both beam momentum settings. The minimum bin size is then 260 MeV. The reconstructed slice distributions, highlighting both the binning and slice distributions as a function of energy, are shown for incident slices in Figure~\ref{fig:recoInc} and for interacting slices in Figure~\ref{fig:recoInt}. These distributions include calorimetric-related uncertainties discussed in Section~\ref{sec:uncertainties}.

\section{Results}
\label{sec:results}

\begin{figure*}[h]
    \centering
    \includegraphics[width=0.45\textwidth]{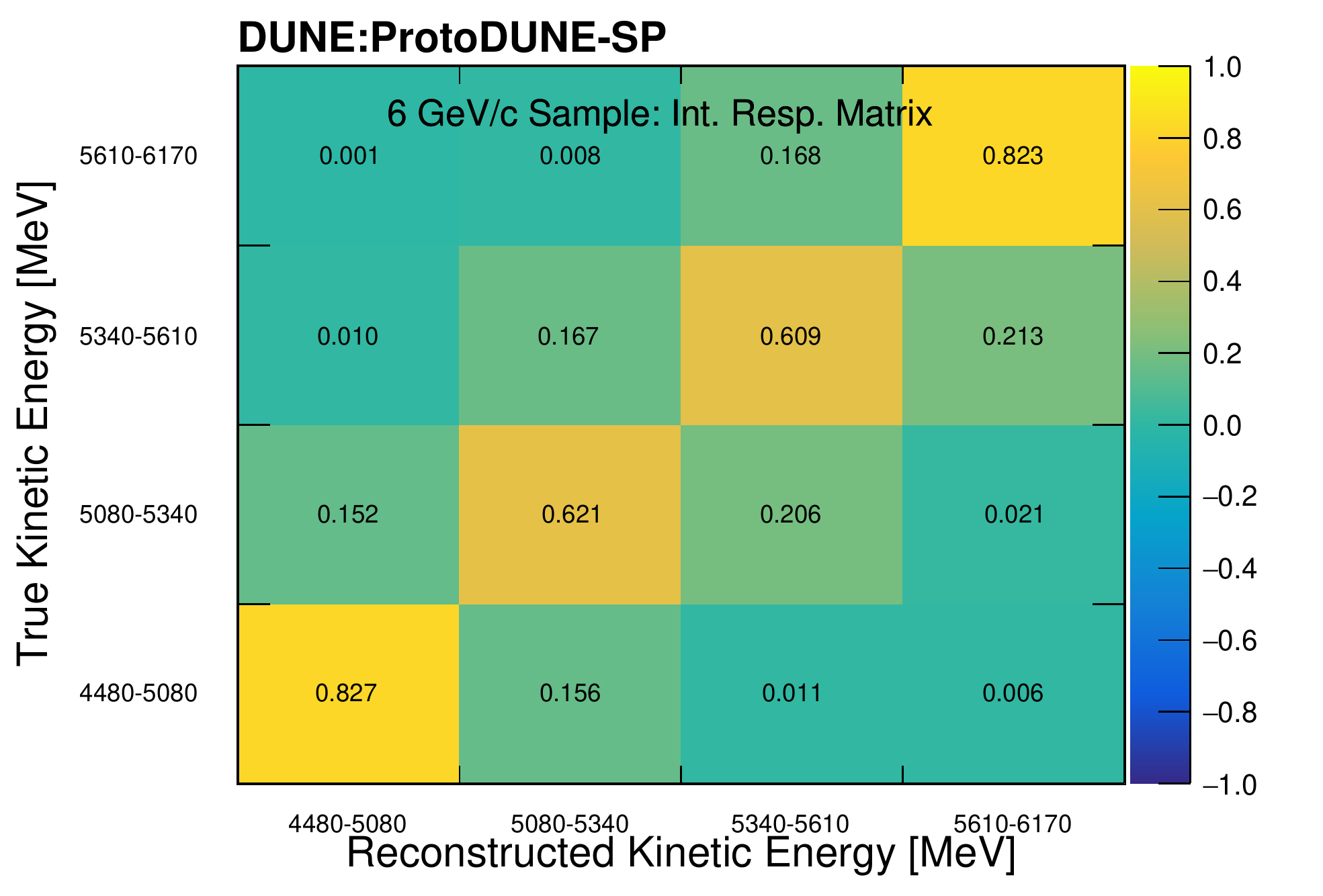}
    \includegraphics[width=0.45\textwidth]{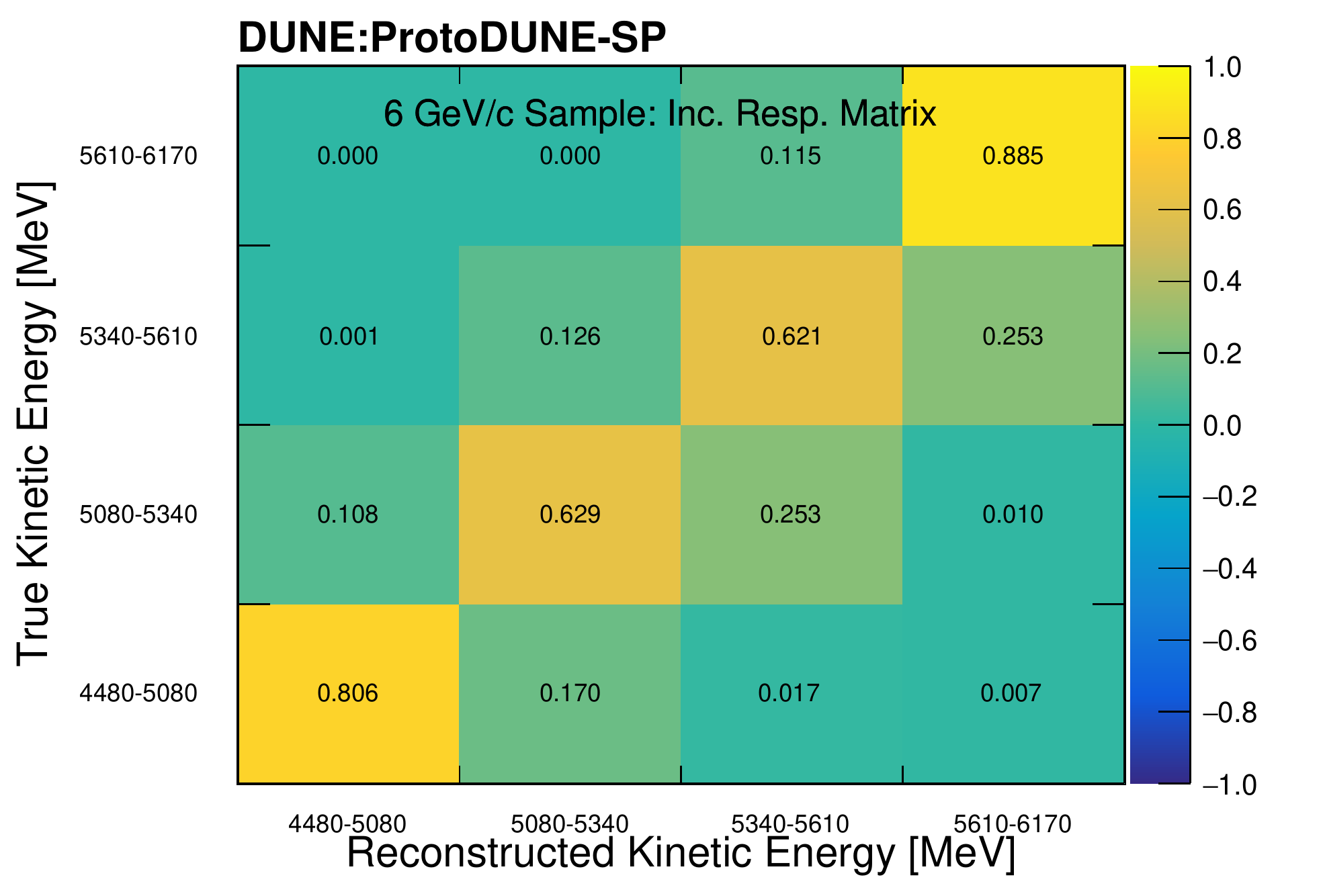}
\caption{Response matrices for the 6 GeV/$c$ simulation sample of the interacting (left) and incident (right) spectra. The entries in the matrices are normalized so that the rows sum to one.}
    \label{fig:responses}
\end{figure*}

\begin{figure*}
    \centering
    \includegraphics[width=0.45\textwidth]{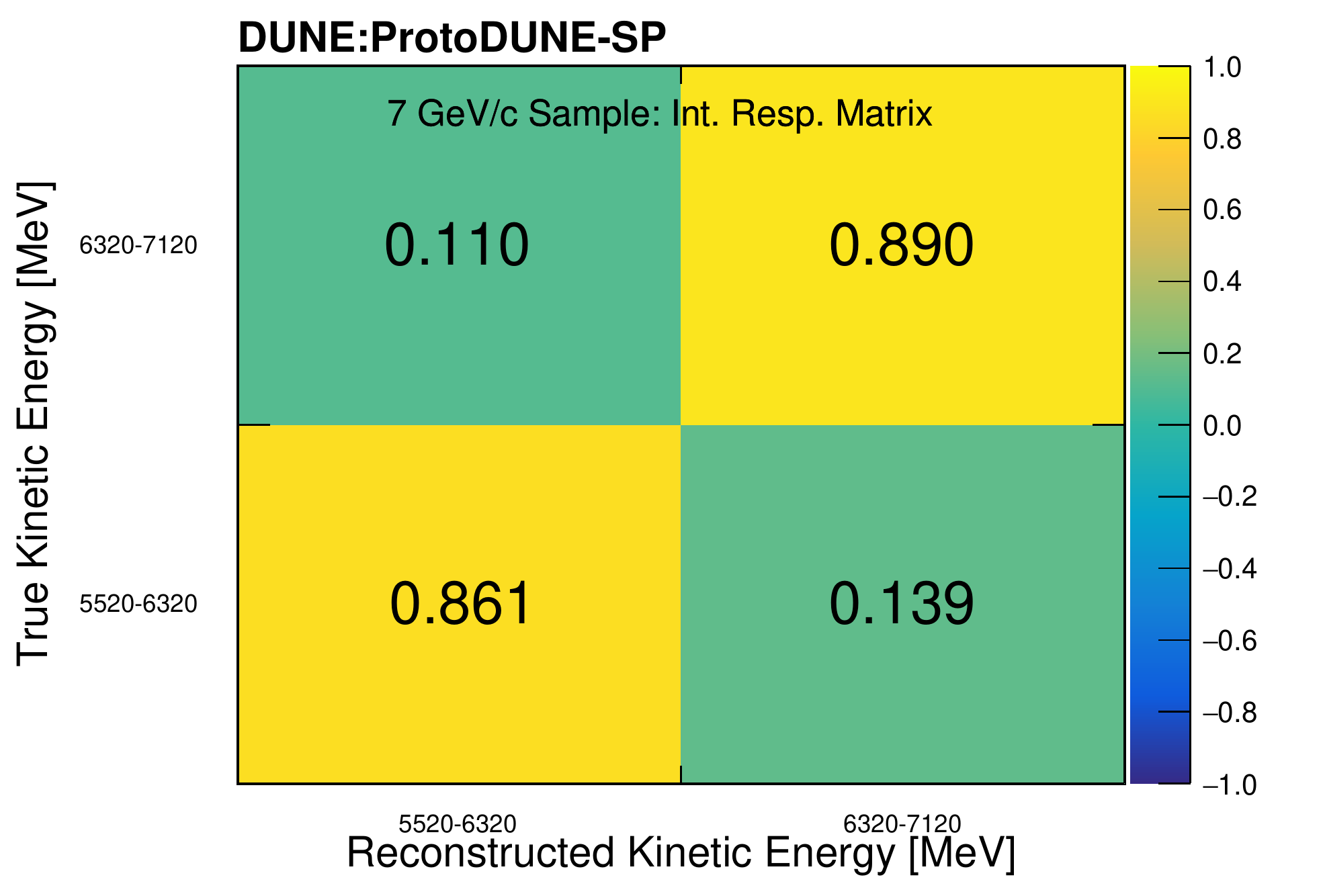}
    \includegraphics[width=0.45\textwidth]{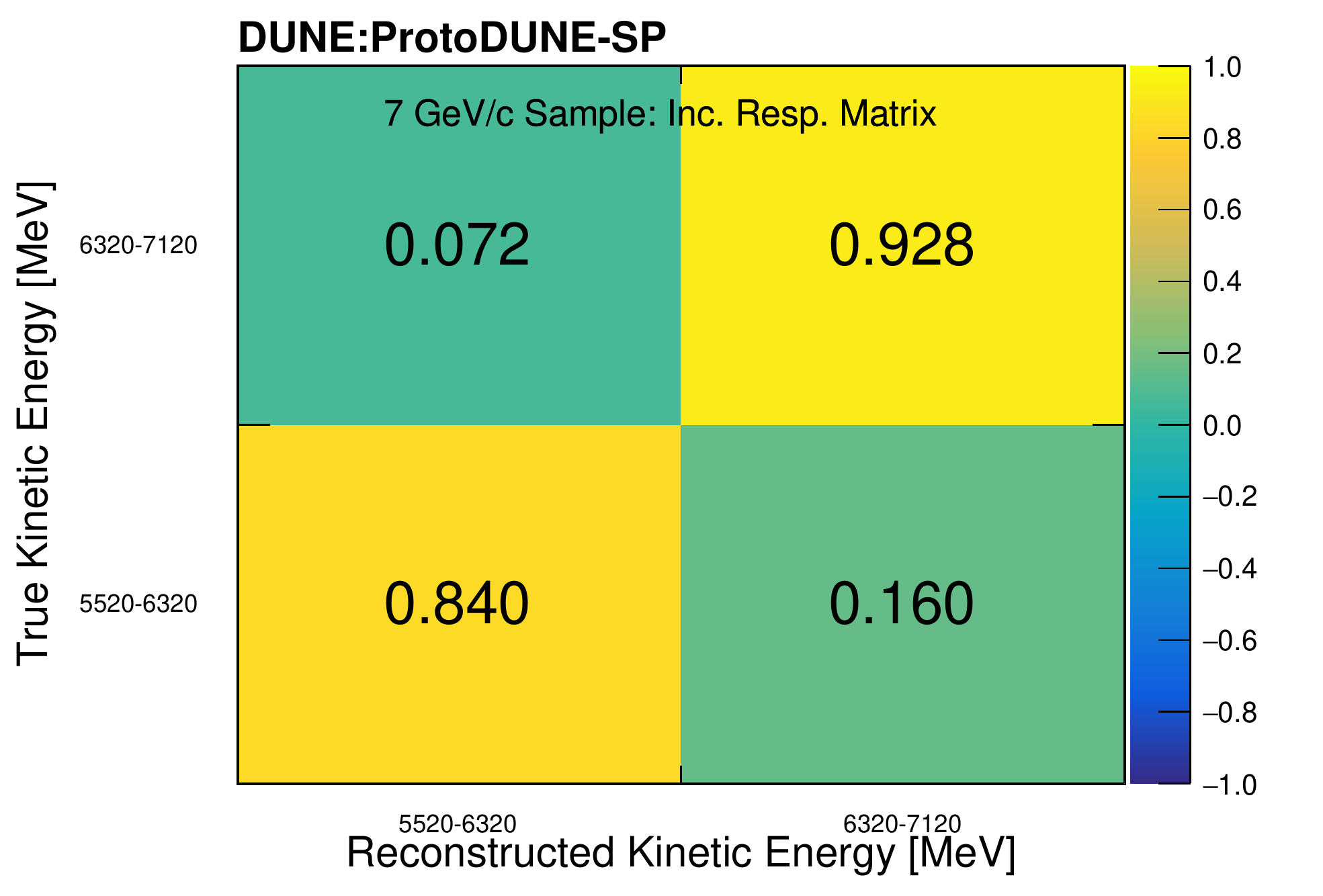}
\caption{Response matrices for the 7 GeV/$c$ simulation sample of the interacting (left) and incident (right) spectra. The entries in the matrices are normalized so that the rows sum to one.}
    \label{fig:responses7GeV}
\end{figure*}

 The kinetic energy distributions for all kaons, and for interacting kaons, are separately unfolded using the method of D'Agostini with four iterations~\cite{DAGOSTINI1995487,Richardson,Lucy,RooUnfoldPaper}. The smearing matrices are shown in Figure~\ref{fig:responses} and Figure~\ref{fig:responses7GeV}. Studies were done to test unfolding by altering the regularization, not correcting for bin-to-bin smearing, and changing the background subtraction and efficiency corrections. All had an impact of less than a percent on the average cross section compared to the nominal unfolding process described above. The response matrix is obtained using only 66\% of the simulated data, which was done to use the remaining 33\% as statistically-independent fake data samples for investigating systematic uncertainties.

\begin{figure*}
    \centering
    \includegraphics[width=0.45\textwidth]{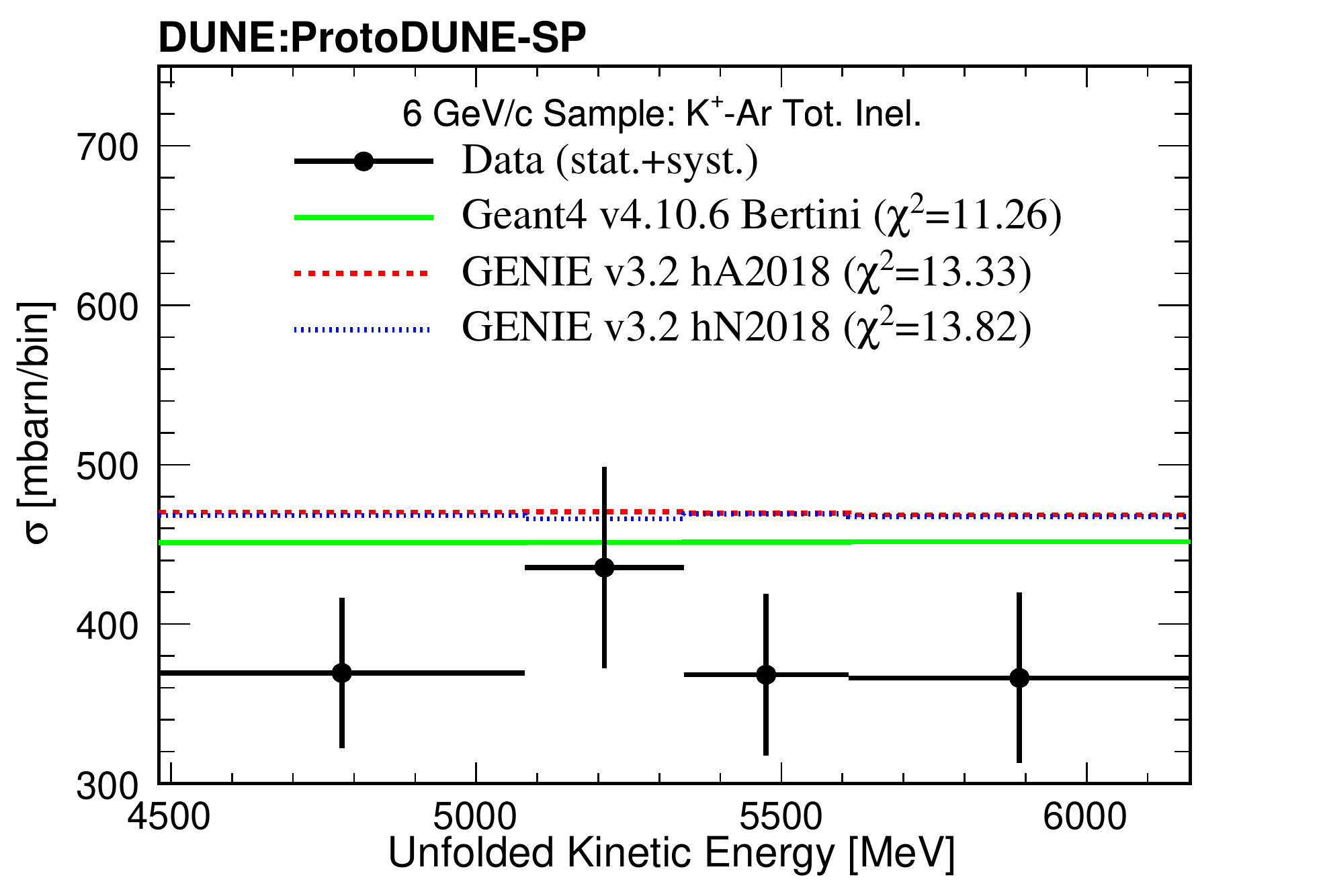}
        \includegraphics[width=0.45\textwidth]{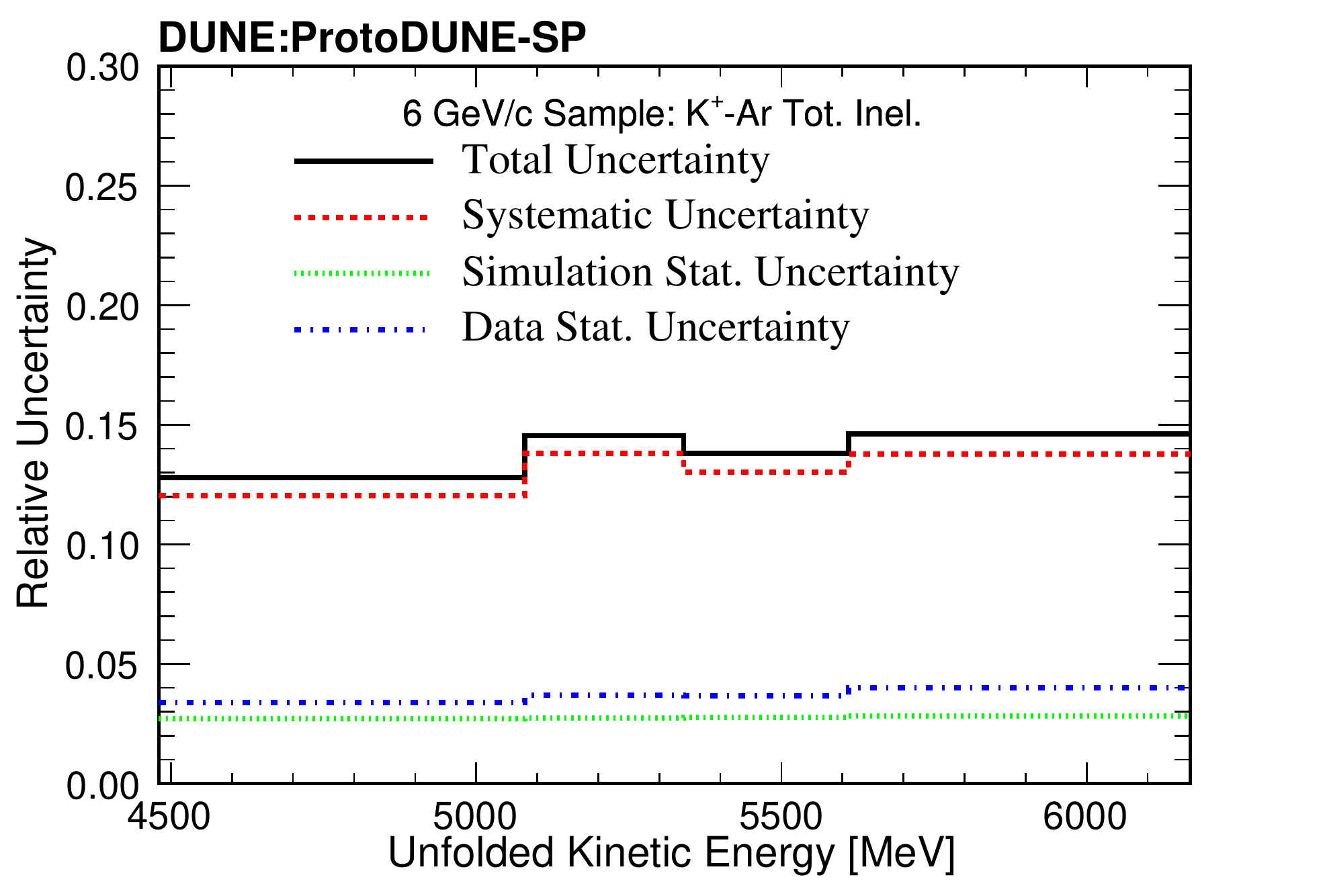}
    \caption{Extracted total inelastic cross section from beam kaons at the momentum setting of 6 GeV/$c$ (left) with comparisons to GENIE v3.2.0 and \textsc{Geant4} ~\cite{GEANT4,geant4IEEE,ALLISON2016186,Andreopoulos:2009rq,Andreopoulos:2015wxa,Tena-Vidal:2021rpu,GENIE2021wox,Dytman}. The relative uncertainties of the measurements are also shown (right). The hA2018 and hN2018 cascade simulations of GENIE provide nearly the same prediction, and their distributions overlap.}
    \label{fig:xsecData}
\end{figure*}

The reconstructed slice spectra, shown in Figure~\ref{fig:recoInc} and Figure~\ref{fig:recoInt}, are unfolded and then used to measure the cross section with Equation~\ref{eqn:xsecthin}. Figure~\ref{fig:xsecData} displays the result for the data of the 6 GeV/$c$ sample with comparisons to predicted cross sections from \textsc{Geant4}, GENIE v3.2.0 hA2018, and GENIE v3.2.0 hN2018~\cite{GEANT4,geant4IEEE,ALLISON2016186,Andreopoulos:2009rq,Andreopoulos:2015wxa,Tena-Vidal:2021rpu,GENIE2021wox,Dytman}. GENIE calculates the total cross section using data and partial wave analysis~\cite{gwu}. It simulates interactions with either an empirical model (hA2018) or a fully simulated cascade (hN2018)~\cite{Andreopoulos:2015wxa, Dytman}. \textsc{Geant4} applies alterations of the base model cross section using data sets included in the Particle Data Group summary cross-section measurements~\cite{pdg2022}. The reduced chi-squared statistics between these models over four bins are 11.26 for \textsc{Geant4} and 13.33 for GENIE v3.2.0 hA2018. The 6 GeV/$c$ data sample flux-averaged cross section is measured at 380$\pm$26 mbarns. Table~\ref{tab:final6GeV} shows the final results with the breakdown of the uncertainties applied.

\begin{table}
\setlength\extrarowheight{1.2pt}
        \caption{Total inelastic positively-charged kaon cross section with uncertainties for data from the 6 GeV/$c$ momentum setting beam. The total uncertainties ($\mathrm{\delta_{tot}}$) are broken down into the systematic uncertainty ($\mathrm{\delta_{syst}}$), statistical uncertainty from limited data statistics ($\mathrm{\delta^{Data}_{stat}}$), and statistical uncertainty from limited simulation statistics ($\mathrm{\delta^{Sim}_{stat}}$). All units for the cross section and uncertainties are in millibarns.}
    \centering
    \begin{tabular}{|c|c|c|c|c|c|} \hline 
     Energy bin (MeV) & $\mathrm{\sigma_{inel}}$ (mbarns) & $\mathrm{\delta_{tot}}$  & $\mathrm{\delta_{syst}}$  & $\mathrm{\delta^{Data}_{stat}}$  & $\mathrm{\delta^{Sim}_{stat}}$   \\ \hline 
4480-5080 & 369 & 47 & 44 & 13 & 10  \\ \hline 
5080-5340 & 435 & 63 & 60 & 16 & 12  \\ \hline 
5340-5610 & 368 & 51 & 48 & 13 & 10  \\ \hline 
5610-6170 & 366 & 54 & 50 & 15 & 10  \\ \hline 
    \end{tabular}
    \label{tab:final6GeV}
\end{table}

\begin{figure*}
    \centering
    \includegraphics[width=0.45\textwidth]{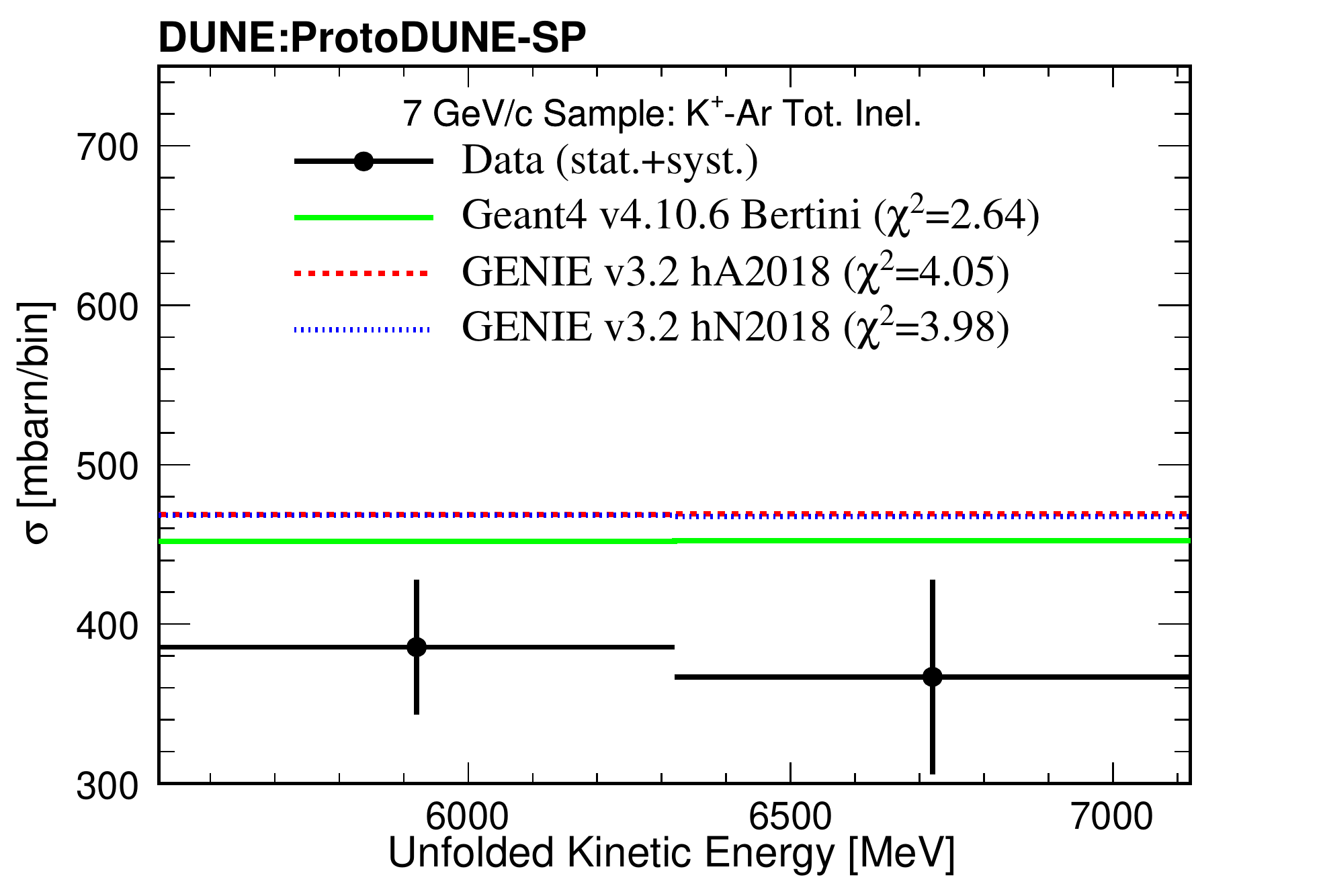}
        \includegraphics[width=0.45\textwidth]{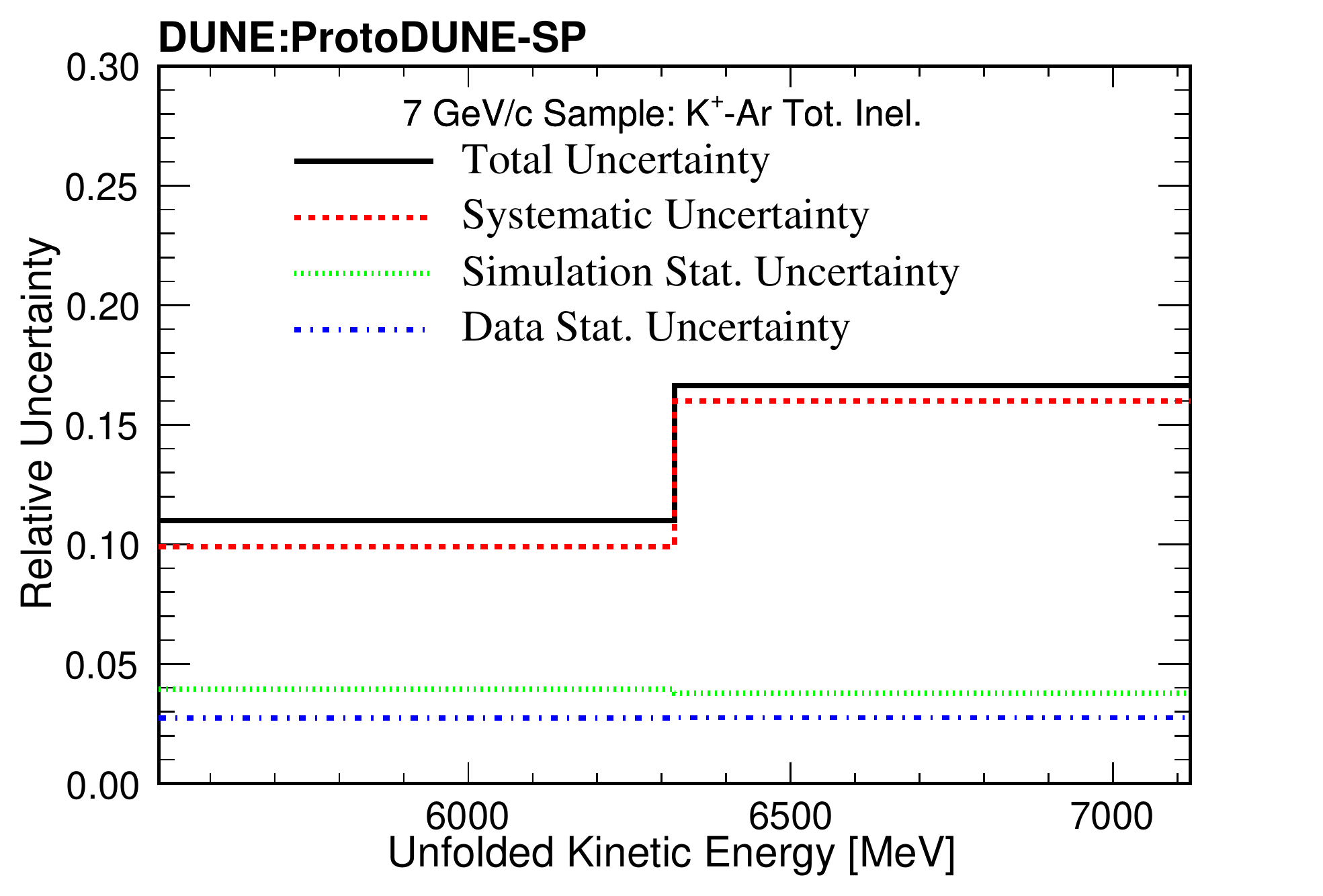}
    \caption{Extracted total inelastic cross section from beam kaons at the momentum setting of 7 GeV/$c$ (left) with comparisons to GENIE v3.2.0 and \textsc{Geant4} ~\cite{GEANT4,geant4IEEE,ALLISON2016186,Andreopoulos:2009rq,Andreopoulos:2015wxa,Tena-Vidal:2021rpu,GENIE2021wox,Dytman}. The relative uncertainties of the measurements are also shown (right). The hA2018 and hN2018 cascade simulations of GENIE provide nearly the same prediction, and their distributions overlap.}
    \label{fig:xsec7GeVcData}
\end{figure*}

Figure~\ref{fig:xsec7GeVcData} presents the cross section measured with data at the 7 GeV/$c$ beam setting. The reduced chi-square statistic measured divided by the number of bins is 2.64/2 bins for \textsc{Geant4} and 4.05/2 bins for GENIE v3.2 hA2018. Table~\ref{tab:final7GeV} displays the final result with uncertainties broken down by category. The flux-averaged cross section is 379$\pm$35 mbarns for the 7 GeV/$c$ sample. Encouragingly, the bin whose energy range overlaps with the 6 GeV/$c$ sample has a similar measured cross section which is within uncertainties.

\begin{table}
\setlength\extrarowheight{1.2pt}
        \caption{Total inelastic positively-charged kaon cross section with uncertainties for data from the 7 GeV/$c$ momentum setting beam. The total uncertainties ($\mathrm{\delta_{tot}}$) are broken down into the systematic uncertainty ($\mathrm{\delta_{syst}}$), statistical uncertainty from limited data statistics ($\mathrm{\delta^{Data}_{stat}}$), and statistical uncertainty from limited simulation statistics ($\mathrm{\delta^{Sim}_{stat}}$). All units for the cross section and uncertainties are in millibarns.}
    \centering
    \begin{tabular}{|c|c|c|c|c|c|} \hline 
     Energy bin (MeV) & $\mathrm{\sigma_{inel}}$ (mbarns) & $\mathrm{\delta_{tot}}$  & $\mathrm{\delta_{syst}}$  & $\mathrm{\delta^{Data}_{stat}}$  & $\mathrm{\delta^{Sim}_{stat}}$   \\ \hline 
5520-6320 & 386 & 42 & 38 & 11 & 15
 \\ \hline 
6320-7120 & 367 & 61 & 59 & 10 & 14
 \\ \hline 
    \end{tabular}
    \label{tab:final7GeV}
\end{table}

\section{Treatment of Uncertainties}
\label{sec:uncertainties}

Uncertainties are propagated by randomly sampling statistical and systematic parameters 1000 times within their a priori uncertainties~\cite{Roe}.
The impact of the statistical uncertainty on the individual kinetic energy bins of the incident and interaction spectra is assessed by randomly Poisson-fluctuating the number of entries in each bin independently according to the measured counts. It is done this way as the statistical uncertainty of the cross section is not directly proportional to the statistical uncertainty of interaction points given that the interaction and incident distributions are inside a logarithm, as seen in Equation~\ref{eqn:xsecthin}. However, there are enough statistics in the incident and interacting distributions to assume both are Poisson-distributed and uncorrelated; therefore, doing many independent fluctuations of each bin can acquire the statistical uncertainty on the cross section by remeasuring the cross section with each Poisson-fluctuated sample. The resulting uncertainty on the measured cross section is approximately 2.7\%, according to Table~\ref{tab:final6GeV} and Table~\ref{tab:final7GeV}.

The finite statistics of the simulation sample primarily impacts the analysis via the background subtraction, unsmearing, and efficiency corrections. This effect is propagated by varying the number of counts in each kinetic energy bin for the backgrounds, inefficiencies, and within the response matrix. Values from sampling a Poisson distribution are used to regenerate the response matrices and affiliated corrections, with each bin treated as independent from the others. As statistics in the incident slices are significantly larger than those of the interacting slices, by a factor of around 200 according to Figure~\ref{fig:recoInc}, this uncertainty is only applied to the response matrix and affiliated corrections for interacting slices.

The systematic uncertainties considered in this analysis are related to the simulation of the beamline and beamline instrumentation, the TPC response, the hadron transport model, and instances in which modeling and reconstruction of the TPC data are ill-posed. The results from unfolding with the new response matrices are used to address the total impact of the systematic effects on the analysis.

 The systematic term associated with the beamline momentum scale arises from uncertainties in terms of the position of the fiber monitors and the magnetic field strength in the beamline instrumentation. This value was calculated as 1.2\%~\cite{CalcuttThesis} of the beam particle energy. Therefore, the energy of the beam kaon is fluctuated according to a Gaussian distribution with width of 1.2\%, resulting in an approximately 2\% uncertainty on the measured cross section.

Furthermore, the particle itself can ``scrape'' against material upon entering the liquid argon, losing energy in the process. These beam ``scrapers'' should appear in selections if their position is greater than 1.5 times the radius of the beam away from the beam center ($r_{\rm{beam}}$). There are 3.15 times more selected events in data that exceed this 1.5$r_{\rm{beam}}$ metric than in simulation. Therefore, the systematic uncertainty treatment alters the frequency of the beam ``scrapers'' with a central value weight of 3.15 and a standard deviation of 2.15 to address the difference between data and simulation. The weight upscales simulation events whereby the beamline instrumentation system momentum and truth information momentum at the TPC differ by over 200 MeV, the estimated minimal energy lost for a beam ``scraper.''

A 3\% calorimetric uncertainty from the TPC is assumed in the energy determination. This value is taken from evaluations of the calorimetry calibration uncertainty~\cite{Diurba_thesis}. That study observed the spread in charge calibration results from subsamples of cosmic-ray muons separated by their trajectories in the detector, based on a similar analysis from MicroBooNE~\cite{Adams_2020}. Although the ProtoDUNE-SP study measured 2\% deviations in calibration values, a 3\% uncertainty is applied to address time-dependent fluctuations in the spread of calibration results. 


The space charge effect, as discussed in Section~\ref{sec:reco}, may alter the reconstructed positions of particles. The fiducial volume is defined to reduce the impact of the space charge effect on the track reconstruction efficiency. Mismodeling of the space charge distribution in the TPC can be effectively treated as shifts in the boundaries of the fiducial volume. The impact of the uncertainty of the space charge modeling is estimated using the spread in the mean distortion at the surfaces of the detector over time. The spread, and subsequently the uncertainty, was measured to be 8\%.

The systematic uncertainty in this analysis arising from mismodeling of charged kaon scattering is assessed by using the Geant4Reweight package~\cite{Calcutt_2021} to reweight events based on the total signal cross section, which intends to probe how the underlying simulated cross section impacts the background subtraction and efficiency corrections. The total inelastic cross section was varied by 20\%. Additionally, due to differences in vertexing observed based on the final state, the kaon multiplicity in the final state of the interaction is reweighted by 20\%, and that weighting is done in a manner to hold the total cross section constant. The impact of all these modeling uncertainties is 2-6\% on the cross section per bin.

The impact of mismodeling the effect of the electron diverter is determined by changing the frequency of these events occurring in the incident response matrices. The following uncertainty reweights the number of tracks broken between 220 and 234 cm inside the detector. It only applies to tracks that do not end within the fiducial volume, which means this weight only impacts the incident slice spectrum. The uncertainty on this effect is set to 100\% as the rate at which the electron diverter breaks reconstructed tracks is not well simulated and overpredicts the number of broken tracks, as seen in Figure~\ref{fig:kaonNoSelTrkLen}.

\begin{figure}
    \centering
    \includegraphics[width=0.45\textwidth]{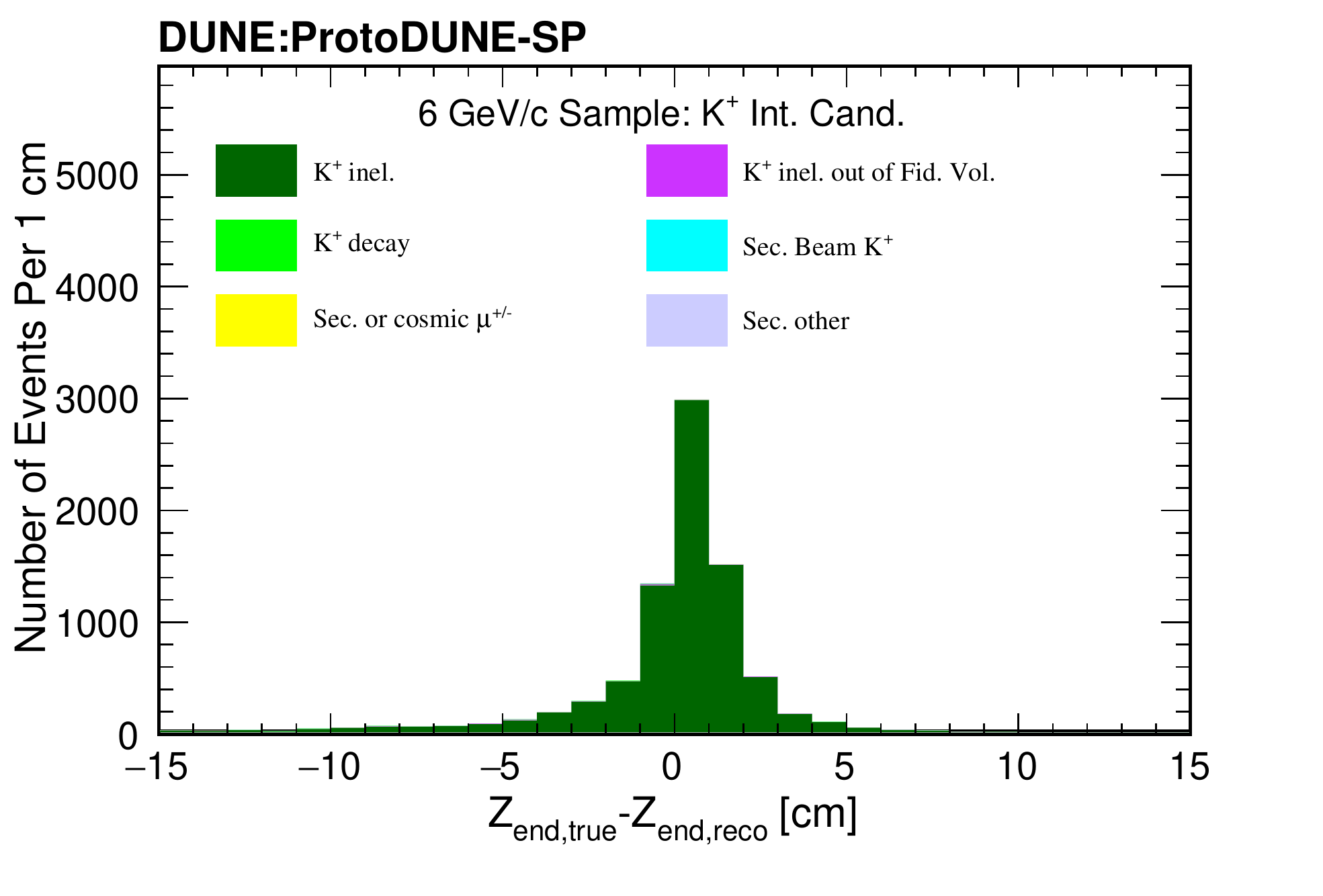}
    \caption{Difference in the endpoint along the detector length (Z) between the truth-level information and the calibrated reconstructed information for the 6 GeV/$c$ simulation sample. The mean offset measured in the 6 GeV/$c$ simulation sample is 0.539 cm with a standard deviation of 1.231 cm using a Gaussian fit.}
    \label{fig:trueVertex}
\end{figure}

While Pandora employs various algorithms to find the endpoint of the beam particle, they may miss the vertex, as Figure~\ref{fig:trueVertex} shows that the endpoint of the reconstructed track may not exactly be the true endpoint. These events may still have kaon inelastic scatters in the fiducial volume. However, these events may underestimate or overestimate the number of incident slices of the beam particle, biasing the results. These tracks are called either broken or extended tracks. An additional uncertainty term was introduced to address the miscounting of incident slices from inaccurate vertex positions, changing the ``flux'' in a \textit{thin slice} measurement. The uncertainty alters the frequency of reconstructed tracks with displacements larger than 5 cm away from the true endpoint along the axis of the detector length. A 100\% uncertainty on their frequencies in simulation is assumed for both broken and extended tracks, as data-driven constraints cannot be provided on vertexing. 

The fiducial volume is chosen to minimize the impact of the tracking inefficiency. However, the TPC track reconstruction may still have discrepancies in performance between data and simulation not covered by the space charge effect systematic uncertainty. Therefore, an uncertainty of 6\% is applied to events without a TPC track, which is a conservative value from the measurements of the efficiency for selecting a beam particle in Reference~\cite{pandoraProtoDUNE}.

Table~\ref{tab:1DPosShifts} shows the $\pm1\sigma$ shifts for data from the 6 GeV/$c$ beamline setting. Table~\ref{tab:1DPosShifts7GeV} reveals the same shifts for data from the 7 GeV/$c$ beamline setting. The dominant uncertainties are the extended track uncertainty and the uncertainty on the \textsc{Geant4} model used in the simulation. The former can be improved with in-depth vertexing studies on how the reconstruction delineates vertices and how much energy is required to create a vertex or stitch the parent and secondary. The latter can be improved with the reduction of the background of secondary kaons reconstructed as the beam particle as the uncertainty alters the frequency of all kaons in simulation, even those in events where the background is selected by the TPC track reconstruction. There is currently no known way to reduce the TPC track selection choosing a secondary kaons, as the $dE/dx$ would be nearly identical to that of beam kaons.

\begin{table*}[h]
\setlength\extrarowheight{3pt}
    \centering
        \caption{Percent deviations from central-value data results by throwing positive one and negative one standard deviation shifts of the uncertainty parameters for the 6 GeV/$c$ sample.}    
        \begin{tabular}{|c|c|c|c|c|} \hline 
    Uncertainty Source ($^{+1\sigma}_{-1\sigma}$) & 4480-5080 MeV (\%) & 5080-5340 MeV (\%) & 5340-5610 MeV (\%) & 5610-6170 MeV (\%) \\ \hline 
    Beam modeling & $^{-1.79}_{\hspace{2mm} 1.58}$ & $^{\hspace{2mm} 2.50}_{-3.89}$ & $^{-0.74}_{\hspace{2mm} 1.71}$ & $^{\hspace{2mm} 4.01}_{\hspace{2mm} 0.51}$ \\ \hline
$dE/dx$ calibration & $^{\hspace{2mm} 
 0.94}_{\hspace{2mm}  1.59}$ & $^{-0.71}_{-0.76}$ & $^{-0.96}_{-1.69}$ & $^{\hspace{2mm} 1.92}_{\hspace{2mm}  1.66}$ \\ \hline
Space charge effect & $^{\hspace{2mm} 1.28}_{\hspace{2mm} 1.92}$ & $^{-1.18}_{\hspace{2mm}  0.76}$ & $^{-2.04}_{\hspace{2mm} 0.28}$ & $^{\hspace{2mm} 2.05}_{\hspace{2mm} 4.42}$ \\ \hline
\textsc{Geant4} modeling & $^{\hspace{2mm} 6.84}_{-4.60}$ & $^{\hspace{2mm} 3.72}_{-5.60}$ & $^{\hspace{2mm} 1.98}_{-5.16}$ & $^{\hspace{2mm} 4.32}_{-0.64}$ \\ \hline
Electron diverter effect & $^{\hspace{2mm} 6.54}_{-1.24}$ &  $^{\hspace{2mm} 3.11}_{-2.68}$ & $^{\hspace{2mm} 1.64}_{-2.73}$ & $^{\hspace{2mm} 2.42}_{\hspace{2mm} 3.43}$ \\ \hline
Vertex identification & $^{\hspace{2mm} 8.55}_{-6.25}$ & $^{\hspace{2mm} 9.37}_{-10.57}$ & $^{\hspace{2mm} 7.93}_{\hspace{-1.2mm} -10.18}$ & $^{\hspace{1mm} 13.44}_{-8.28}$ \\ \hline
Events without a track & $^{\hspace{2mm} 1.61}_{\hspace{2mm} 1.22}$ & $^{-0.29}_{-1.40}$ & $^{-1.05}_{-1.83}$ & $^{\hspace{2mm} 2.70}_{\hspace{2mm} 1.27}$ \\ \hline
Simulation statistics & $^{-0.90}_{\hspace{2mm}0.89}$ & $^{-1.81}_{\hspace{2mm}2.12}$ & $^{-1.54}_{\hspace{2mm}1.76}$ & $^{-2.27}_{\hspace{2mm}2.48}$ \\ \hline
Data statistics & $^{\hspace{2mm}2.65}_{-2.27}$ & $^{-4.80}_{-9.77}$ & $^{\hspace{2mm}5.35}_{-0.06}$ & $^{\hspace{2mm}6.58}_{-0.56}$ \\ \hline
All Uncertainties & $^{\hspace{1mm}13.38}_{\hspace{2mm}8.82}$ & $^{\hspace{1mm}12.08}_{\hspace{1mm}16.38}$ & $^{\hspace{1mm}10.35}_{\hspace{1mm}12.25}$ & $^{\hspace{1mm}16.88}_{\hspace{1mm}10.56}$ \\ \hline

    \end{tabular}
    \label{tab:1DPosShifts}
\end{table*}

\begin{table*}[h]
\setlength\extrarowheight{3pt}

    \centering
        \caption{Percent deviations from central-value data results by throwing positive one and negative one standard deviation shifts of the uncertainty parameters for the 7 GeV/$c$ sample.}
    \begin{tabular}{|c|c|c|c|c|} \hline 
    Uncertainty Source ($^{+1\sigma}_{-1\sigma}$) &  5520-6320 MeV (\%)  & 6320-7120 MeV (\%)  \\ \hline 
    Beam modeling & $^{-2.38}_{-3.44}$ & $^{ \hspace{2mm} 2.16}_{-0.10}$ \\ \hline
$dE/dx$ calibration & $^{\hspace{2mm} 0.69}_{-0.61}$ & $^{ \hspace{2mm} 0.18}_{ \hspace{2mm} 1.41}$ \\ \hline
Space charge effect & $^{-0.06}_{\hspace{2mm} 0.85}$ & $^{\hspace{2mm} 0.07}_{ \hspace{2mm} 2.76}$ \\ \hline
\textsc{Geant4} modeling & $^{\hspace{2mm} 4.12}_{-4.23}$ & $^{\hspace{2mm} 2.57}_{-1.05}$ \\ \hline
Electron diverter effect & $^{\hspace{2mm} 3.77}_{-3.46}$ &  $^{-1.69}_{\hspace{2mm} 3.46}$ \\ \hline
Vertex identification & $^{\hspace{2mm} 5.94}_{-7.24}$ & $^{\hspace{1mm} 14.76}_{\hspace{-1.2mm} -12.00}$ \\ \hline
Events without a track & $^{\hspace{2mm} 0.45}_{-0.56}$ & $^{\hspace{2mm} 1.50}_{-0.20}$ \\ \hline
Simulation statistics & $^{-1.87}_{\hspace{2mm}2.04}$ & $^{-2.59}_{\hspace{2mm}3.05}$ \\ \hline 
Data statistics & $^{-0.79}_{-5.03}$ & $^{\hspace{2mm}3.57}_{-1.46}$ \\ \hline 
All Uncertainties & $^{\hspace{2mm}8.78}_{\hspace{1mm} 11.18}$ & $^{\hspace{1mm}15.93}_{\hspace{1mm}13.34}$ \\ \hline

    \end{tabular}
    \label{tab:1DPosShifts7GeV}
\end{table*}

\section{Conclusions}
\label{sec:conclusions}

This paper describes a measurement of the total inelastic cross section of positively-charged kaons on argon with the ProtoDUNE-SP detector using the \textit{thin slice method}~\cite{elena}. This measurement was done with data taken at the H4-VLE at the CERN Neutrino Platform. A simple event selection achieved a purity of approximately 85-90\% between kinetic energies of 4.5 and  7.0 GeV (Table~\ref{tab:standardStats}). The results, at around 380 mbarns, have a precision of approximately 14\%, according to Table~\ref{tab:final6GeV} and Table~\ref{tab:final7GeV}. The total uncertainty almost entirely comes from systematic uncertainties addressing the modeling of the detector and uncertainties regarding the kaon cross-section model used in \textsc{Geant4}. The measurements translate to \textsc{Geant4} overestimating the cross section by 16\% and GENIE overestimating the cross section by 19\%. For the 6 GeV/$c$ data sample, the reduced chi-square statistic measured is 11.26/4 bins with \textsc{Geant4} and 13.33/4 bins with GENIE hA2018, suggesting tension with both models.

Future studies can utilize the results for tuning kaon reaction scattering models incorporated in various hadron interaction event generators. The upcoming ProtoDUNE Horizontal Drift detector will also exist in the same detector hall with a wire-based readout and can measure similar cross sections with nearly identical methods. It may also run with the beam polarity reversed, allowing for a cross-section analysis of negatively-charged kaons.

\section*{Acknowledgements}

%
%
The ProtoDUNE-SP detector was constructed and operated on the CERN Neutrino Platform.
We gratefully acknowledge the support of the CERN management, and the
CERN EP, BE, TE, EN and IT Departments for NP04/Proto\-DUNE-SP.
%
%
This document was prepared by the DUNE collaboration using the
resources of the Fermi National Accelerator Laboratory 
(Fermilab), a U.S. Department of Energy, Office of Science, 
HEP User Facility. Fermilab is managed by Fermi Research Alliance, 
LLC (FRA), acting under Contract No. DE-AC02-07CH11359.
%
%
This work was supported by
CNPq,
FAPERJ,
FAPEG and 
FAPESP,                         Brazil;
CFI, 
IPP and 
NSERC,                          Canada;
CERN;
M\v{S}MT,                       Czech Republic;
ERDF, 
H2020-EU and 
MSCA,                           European Union;
CNRS/IN2P3 and
CEA,                            France;
INFN,                           Italy;
FCT,                            Portugal;
NRF,                            South Korea;
CAM, 
Fundaci\'{o}n ``La Caixa'',
Junta de Andaluc\'ia-FEDER,
MICINN, and
Xunta de Galicia,               Spain;
SERI and 
SNSF,                           Switzerland;
T\"UB\.ITAK,                    Turkey;
The Royal Society and 
UKRI/STFC,                      United Kingdom;
DOE and 
NSF,                            United States of America.

\appendix
\section{Distributions of the 7 GeV/$c$ Beam Event Selections}\label{sec:appendix}

\begin{figure}[h]
    \centering
    \includegraphics[width=0.45\textwidth]{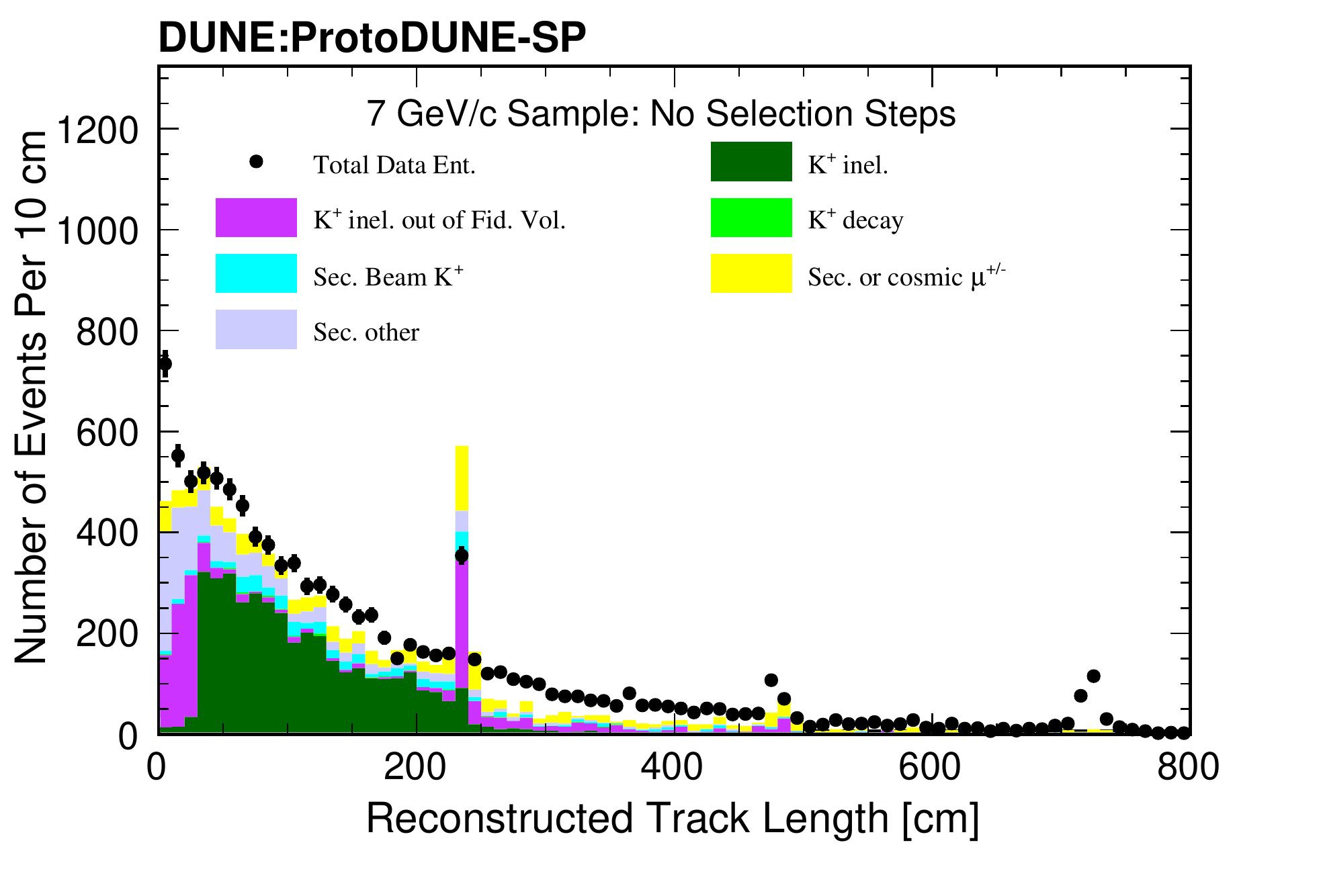}
    \caption{Reconstructed track length for simulation and data without any selection steps for the 7 GeV/$c$ samples. The statistics are scaled according to the data, regardless of if the event had a TPC track. All uncertainties are statistical.}
    \label{fig:kaonNoSelTrkLen7GeV}
\end{figure}

\begin{figure}
    \centering
    \includegraphics[width=0.45\textwidth]{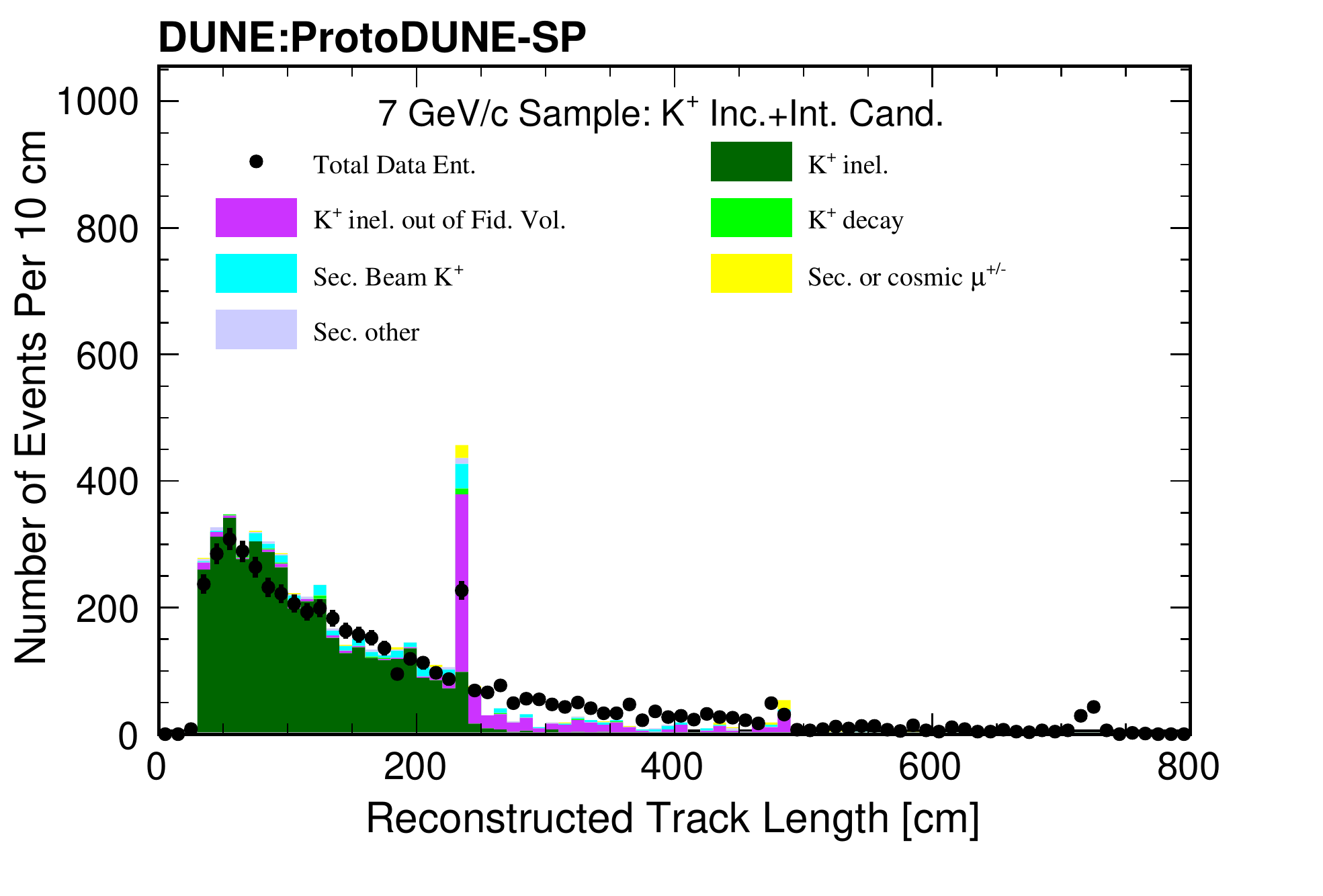}
    \includegraphics[width=0.45\textwidth]{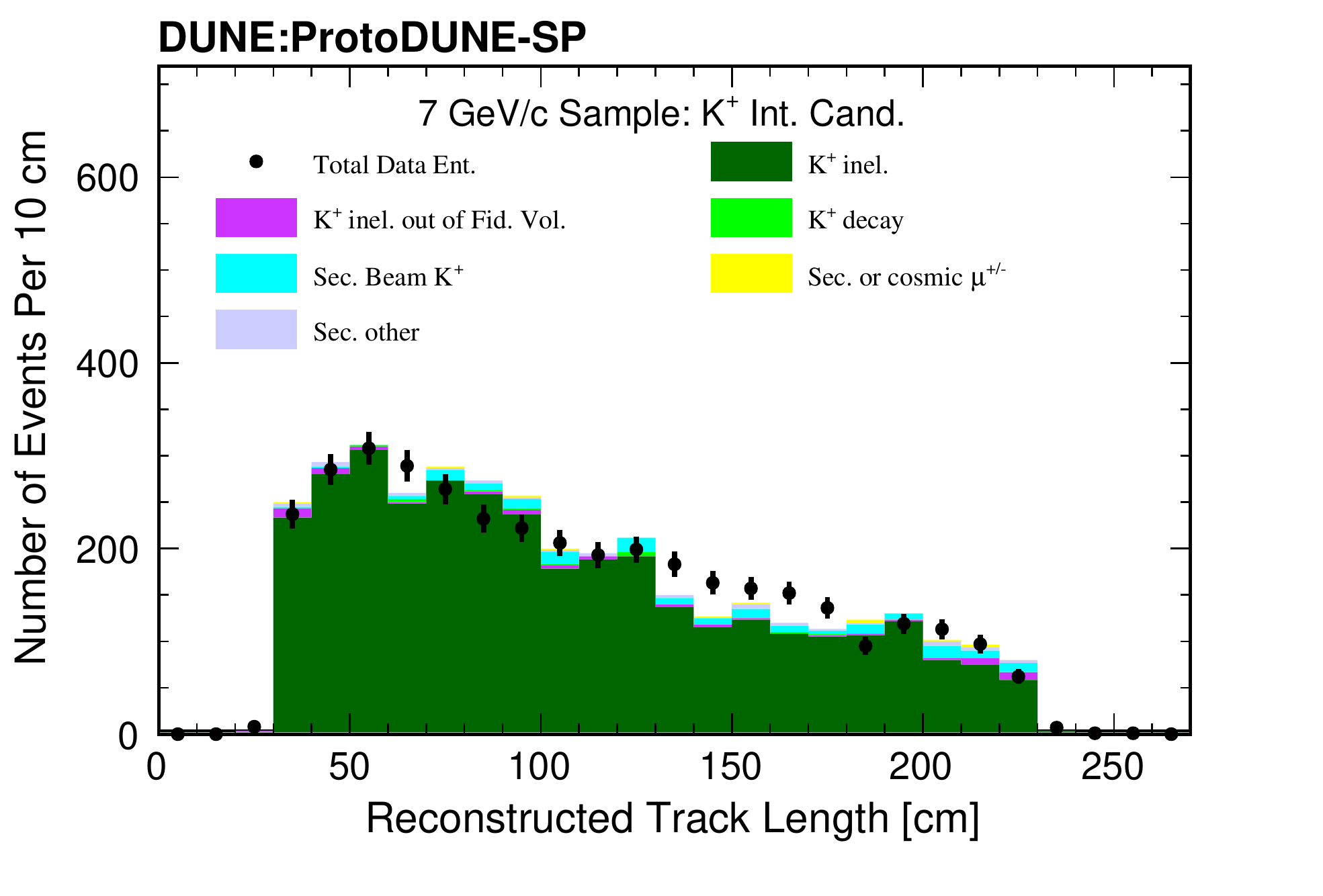}
    \caption{Reconstructed track length for simulation and data for the 7 GeV/$c$ samples both for all selected kaons (top) and only selected kaons with interacting slices in the fiducial volume (bottom). The statistics are scaled according to the data. All uncertainties are statistical.}
    \label{fig:kaonSelAllTrkLen7GeV}
\end{figure}

\begin{figure}
    \centering

        \includegraphics[width=0.45\textwidth]{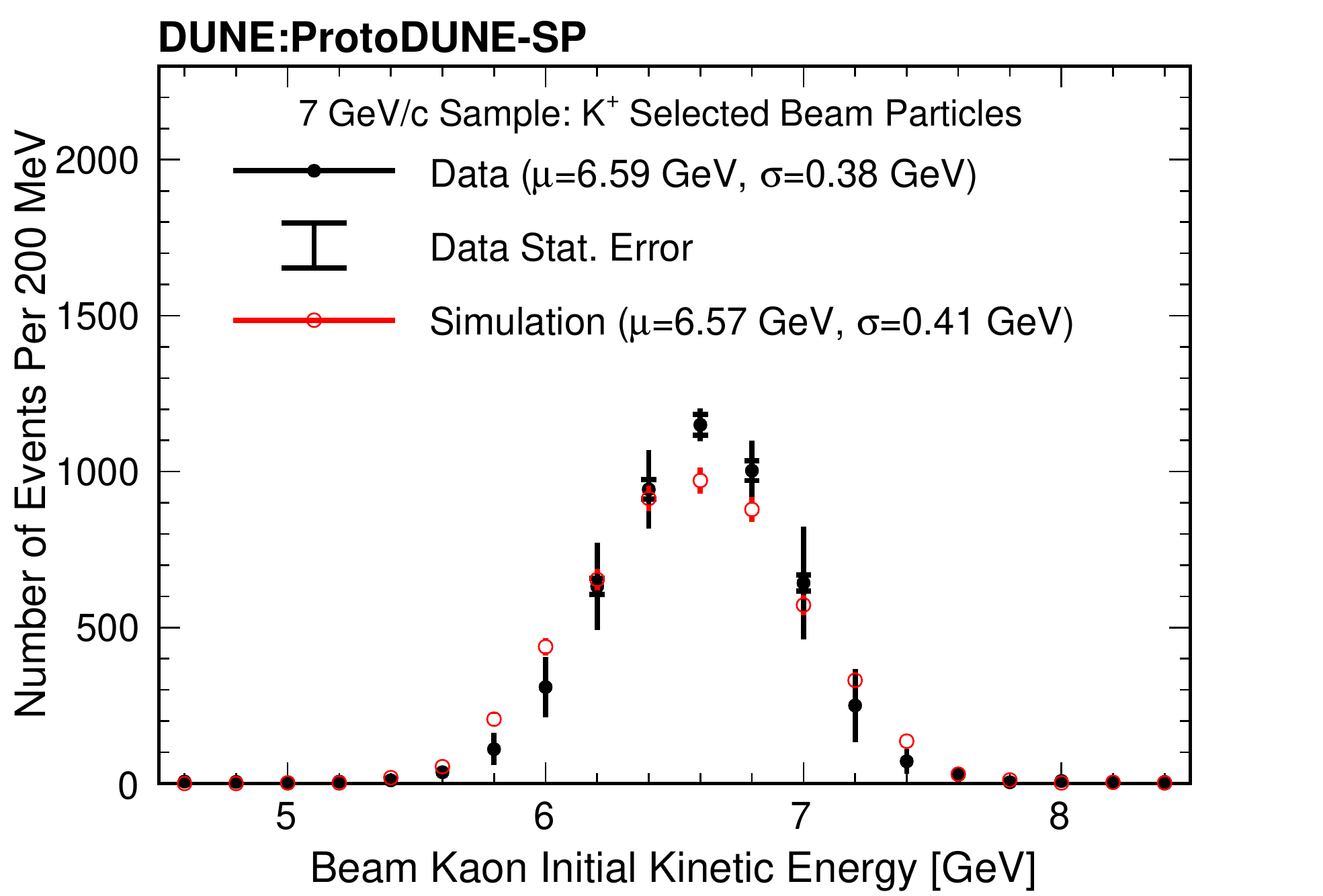}
    \caption{Initial beam particle kinetic energy as measured by the beamline instrumentation for selected kaon candidate tracks for the 7 GeV/$c$ beamline setting. Both systematic and statistical uncertainties are shown.}
    \label{fig:beam7GeV}
\end{figure}

This appendix contains the distributions for the 7 GeV/$c$ samples for the event selection in simulation and data. The distribution of tracks without the event selection are shown in Figure~\ref{fig:kaonNoSelTrkLen7GeV}. The distribution for all selected kaons and selected kaons with interactions in the fiducial volume are shown in Figure~\ref{fig:kaonSelAllTrkLen7GeV}. The initial beamline kinetic energy distributions for all selected beam kaons at this beamline setting, as measured by the beamline instrumentation, is shown in Figure~\ref{fig:beam7GeV}.

\begin{figure}
    \centering
    \includegraphics[width=0.45\textwidth]{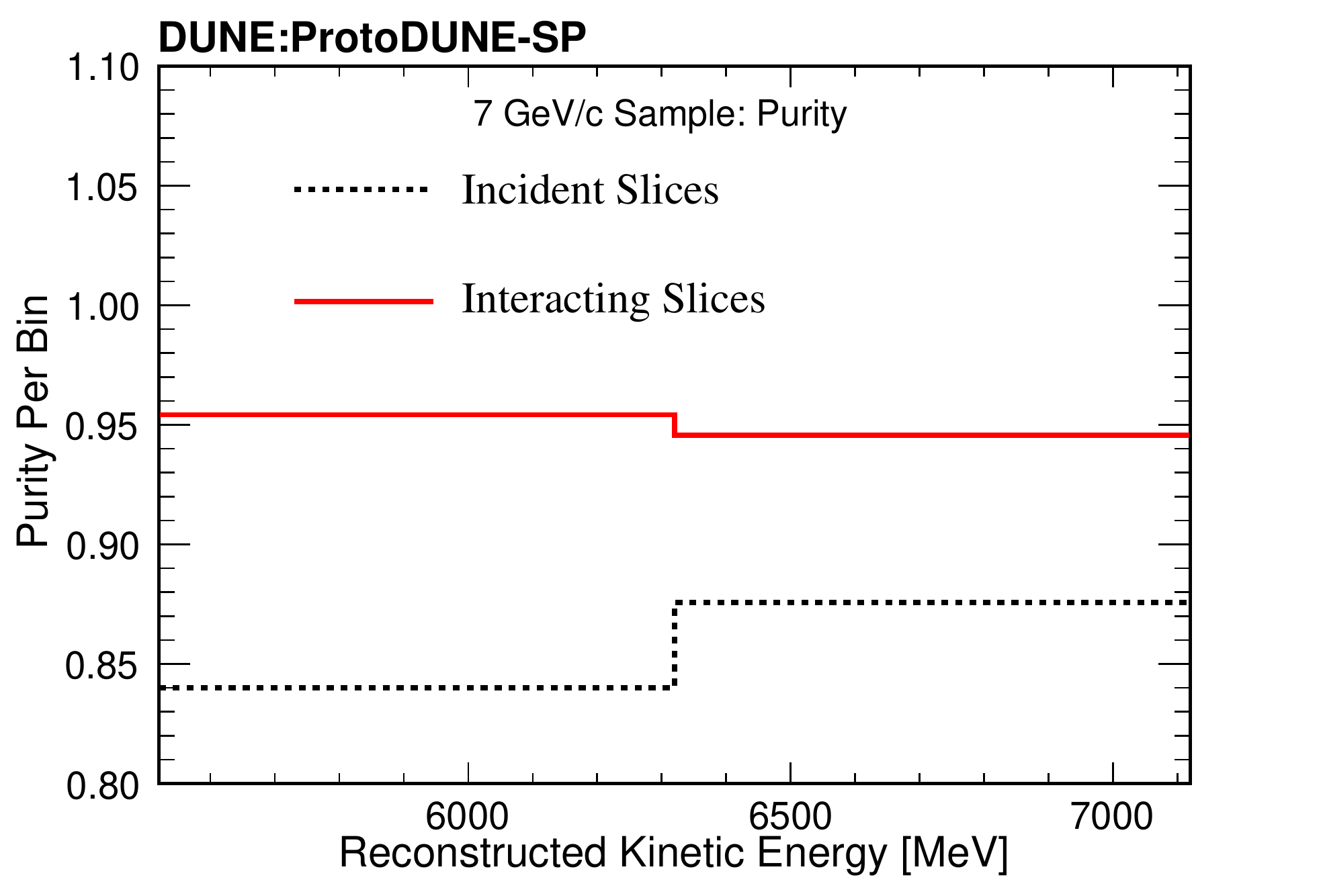}
    \includegraphics[width=0.45\textwidth]{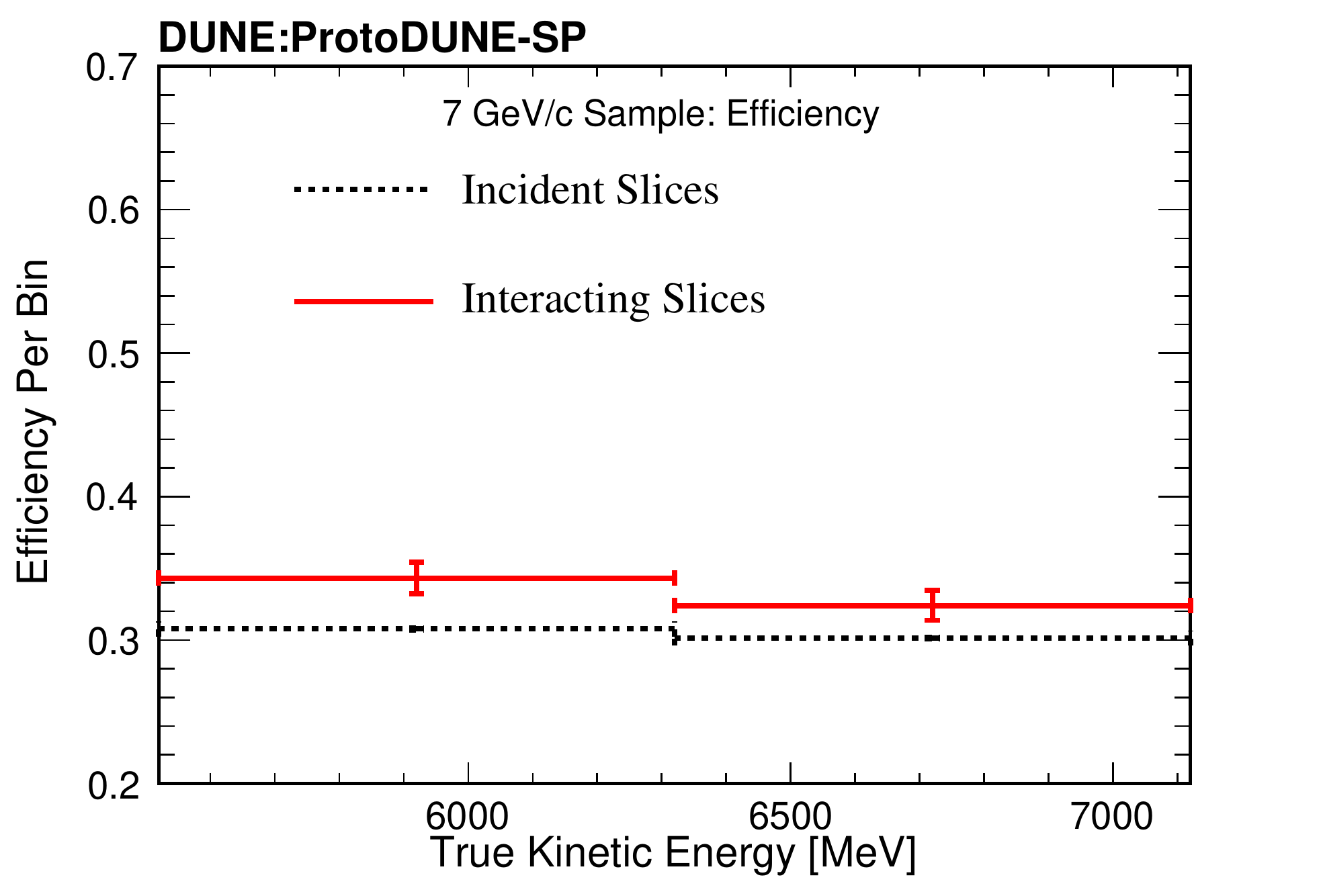}
    \caption{Purity (top) and efficiency (bottom) for each bin using the event selection for the 7 GeV/$c$ simulation sample.}
    \label{fig:purEff7GeV}
\end{figure}

All distributions show agreements with similar distributions in the 6 GeV/$c$ sample, as seen in Figure~\ref{fig:kaonSelAllTrkLen}. Furthermore, similar purities and slightly lower efficiencies in incident and interacting slice distributions can be observed in Figure~\ref{fig:purEff7GeV}.

\FloatBarrier
\bibliography{references}

\end{document}